\newif\ifarxiv
\newif\ifanonymous
\newif\ifincludetodos

\includetodosfalse

\newif\iffigreorg
\figreorgtrue

\arxivfalse

\ifarxiv
    \documentclass[acmsmall,authorversion,nonacm,screen]{acmart}
    \anonymousfalse
\else
    \anonymousfalse
    \ifanonymous
        \documentclass[acmsmall,screen,review,anonymous]{acmart}
    \else
        \documentclass[acmsmall]{acmart}
    \fi
\fi

\ifarxiv\else
\setcopyright{acmlicensed}
\acmDOI{10.1145/3729398}
\acmYear{2025}
\acmJournal{PACMSE}
\acmVolume{2}
\acmNumber{FSE}
\acmArticle{FSE128}
\acmMonth{7}
\received{2024-09-12}
\received[accepted]{2025-04-01}
\fi
\ccsdesc[500]{Computing methodologies~Artificial intelligence}
\ccsdesc[500]{Software and its engineering~Software testing and debugging}

\usepackage[aboveskip=5pt]{caption}
\usepackage{subcaption}

\usepackage{float}
\usepackage{rotating}
\usepackage{makecell}
\usepackage{enumitem}
\usepackage{natbib}
\usepackage{xcolor}
\usepackage{graphicx}
\usepackage{wrapfig}
\usepackage{listings}
\usepackage{siunitx}
\usepackage[utf8]{inputenc}
\usepackage[T1]{fontenc}
\usepackage{microtype}
\usepackage{array}
\usepackage{tabularx}
\usepackage{xstring}

\usepackage{hyperref}
\newcommand{\listingfont}{\ttfamily\scriptsize}

\usepackage{tikz}
\newcommand*\circled[2]{\tikz[baseline=(char.base)]{
    \node[shape=circle,draw,inner sep=2pt, fill=#1] (char) {#2};}}

\usepackage[most]{tcolorbox}
\newtcblisting{colorlisting}[3]{
  beforeafter skip=0pt,
  colback=#1,
  colframe=#1,
  coltitle=black,
  title=\IfStrEq{#2}{}{}{\small\circled{#1}{#2}},
  boxrule=0pt,
  arc=0pt, 
  top=0pt,
  bottom=0pt,
  left=0pt,
  right=0pt,
  listing only,
  listing options={
    basicstyle=\listingfont,
    keywordstyle=\bfseries\color{blue},
    #3
  }
}

\usepackage[ruled,vlined]{algorithm2e}

\newcommand{\codamosa}{\textsc{CodaMosa}}

\ifincludetodos
    \newcommand{\emery}[1]{\textcolor{magenta}{[Emery: #1]}}
    \newcommand{\juan}[1]{\textcolor{blue}{[Juan: #1]}}
    \usepackage[color]{changebar}\cbcolor{red}
\else
    \newcommand{\emery}[1]{}
    \newcommand{\juan}[1]{}
\fi

\newcommand{\punt}[1]{}

\definecolor{mygreen}{rgb}{0,0.6,0}
\definecolor{mygray}{rgb}{0.5,0.5,0.5}
\definecolor{mymauve}{rgb}{0.58,0,0.82}
\ifanonymous
    \newcommand{\redacted}[1]{[redacted]}
\else
    \newcommand{\redacted}[1]{#1}
\fi

\iffalse 
    \newcommand{\coverup}{\textsc{SomeAlias}}
\else
    \newcommand{\coverup}{\textsc{CoverUp}}
\fi

\ifanonymous
    \newcommand{\coverupurl}{\url{https://anonymous.4open.science/r/coverup-62EC}}
    \newcommand{\cleanslateurl}{\url{https://anonymous.4open.science/r/pytest-cleanslate-053D}}
\else
    \newcommand{\coverupurl}{\url{https://github.com/plasma-umass/coverup}}
    \newcommand{\cleanslateurl}{\url{https://github.com/plasma-umass/pytest-cleanslate}}
    \newcommand{\replicationurl}{\url{https://github.com/plasma-umass/coverup-eval}}
\fi

\newcommand{\pct}[1]{\SI{#1}{\%}}
\newcommand{\PCT}[1]{\textbf{\pct{#1}}}

\begin{document}
    \ifarxiv
        \title{\coverup{}: Coverage-Guided LLM-Based Test Generation}
    \else
        \title{\coverup{}: Effective High Coverage Test Generation for Python}
    \fi

    \author{Juan Altmayer Pizzorno}
    \orcid{0000-0002-1891-2919}
    \affiliation{%
        \institution{University of Massachusetts Amherst}
        \city{Amherst}
        \state{MA}
        \country{United States}
    }
    \email{jpizzorno@cs.umass.edu}

    \author{Emery D. Berger}
    \authornote{Work done at the University of Massachusetts Amherst.}
    \orcid{0000-0002-3222-3271}
    \affiliation{%
      \institution{University of Massachusetts Amherst}
      \city{Amherst}
      \country{USA}
    }
    \affiliation{%
      \institution{Amazon Web Services}
      \city{Seattle}
      \country{USA}
    }
    \email{emery@cs.umass.edu}

    \keywords{Test Generation, Regression Testing, Large Language Models, Code Coverage}

    \begin{abstract}
        Testing is an essential part of software development.
Test generation tools attempt to automate the otherwise labor-intensive task of test creation, but generating high-coverage tests remains challenging.
This paper proposes \coverup{}, a novel approach to driving the generation of high-coverage Python regression tests.
\coverup{} combines coverage analysis, code context, and feedback in prompts that iteratively guide the LLM to generate tests that improve line and branch coverage.

We evaluate our prototype \coverup{} implementation across a benchmark of challenging code derived from open-source Python projects and show that \coverup{} substantially improves on the state of the art.
Compared to \codamosa{}, a hybrid search/LLM-based test generator, \coverup{} achieves a per-module median line+branch coverage of 80\% (vs. 47\%).
Compared to MuTAP, a mutation- and LLM-based test generator, \coverup{} achieves an overall line+branch coverage of 89\% (vs. 77\%).
We also demonstrate that \coverup{}'s performance stems not only from the LLM used but from the combined effectiveness of its components.

    \end{abstract}

    \maketitle

    \section{Introduction}
Testing is essential to ensuring software quality, but manually crafting tests can be so labor-intensive that developers choose not to write them~\cite{unit-testing-survey}.
Test generation tools attempt to automate this task, generally assuming that the source code is correct.
By generating tests based on it, they enable \emph{regression testing}, which aims to prevent \emph{future} bugs as the software is modified.
As these tools add tests that cover a wider range of scenarios and execution paths, they can help uncover bugs missed by manually written tests.

This paper proposes \coverup{}, a novel approach to test generation aimed at achieving high coverage.
Our key insight is that large language models (LLMs) can simultaneously reason about code and coverage information.
We leverage this insight in the design of \coverup{}.
\coverup{} incorporates detailed coverage information into prompts customized to the current state of the test suite, focusing the LLM on code that lacks coverage.
Additionally, it provides a \emph{tool function}~\cite{openai-function-calling} that allows the LLM to request additional source code context.
Once the LLM generates a set of tests, \coverup{} executes them and measures the resulting coverage.
If the LLM-generated tests fail to improve coverage or fail to run altogether, \coverup{} continues the dialogue with the LLM, requesting changes to improve coverage or fix the error.
In doing so, \coverup{}, in effect, further refines and clarifies the prompt, resulting in tests that significantly improve coverage.

Our empirical evaluation compares \coverup{} to two state-of-the-art test generation systems: \codamosa~\cite{codamosa}, a hybrid search/LLM-based test generator, and MuTAP~\cite{mutap}, a mutation/LLM-based approach.
We show that \coverup{} substantially improves the state of the art.
Across a benchmark of challenging code derived from open-source Python projects and using OpenAI's GPT-4o LLM, \coverup{} increases coverage on every metric vs. \codamosa{}, achieving per-module median line+branch coverage of 80\% (vs. 47\%) and overall line+branch coverage of 60\% (vs. 45\%).
Compared to MuTAP on one of the benchmark suites it supports, \coverup{} again achieves greater or equal coverage on every metric, with an overall line+branch coverage of 89\% (vs. 77\%).

This paper makes the following contributions:
\begin{itemize}[topsep=0.5\baselineskip, leftmargin=*]
\item It presents \coverup{}, a novel approach that drives the generation of high-coverage Python regression tests via a combination of coverage analysis and large language models;
\item It conducts an empirical analysis of our \coverup{} prototype, showing that it significantly advances the state of the art;
\item It conducts an ablation study, showing that \coverup{}'s approach plays a substantial role in the prototype's performance beyond what can be attributed to the LLM.  
\end{itemize}

    \section{Related Work}
Automated test generation is a well-established field of research.
Among the various proposed methods are specification-based~\cite{korat}, random~\cite{miller-fuzzing, dissecting-afl}, feedback-directed random~\cite{randoop}, symbolic execution guided random (``concolic'')~\cite{dart, cute, pex}, search-based software testing (SBST)~\cite{mutation-driven-generation, evosuite, pynguin, mosa, dynamosa} and transformer-based approaches~\cite{athenatest}.

The success of large language models on various tasks has motivated their application to software engineering; Wang et al.\ survey 102 recent papers using LLMs for software testing~\cite{wang2024survey}.
This section focuses on previous work most closely related to \coverup{}.

\subsubsection*{SBST approaches}
Pynguin~\cite{pynguin} employs a \emph{search-based software testing} (SBST) approach.
Starting from randomly created test cases, SBST employs genetic algorithms to mutate the tests, aiming to increase coverage.
Unfortunately, its search process can get stuck as test mutations repeatedly lead to the same execution paths.
\codamosa~\cite{codamosa} addresses this problem by tracking Pynguin's progress; when it concludes that the search process has stalled, it prompts an LLM for a test.
It then uses that test to re-seed the SBST, allowing it to resume progress.
We empirically compare \coverup{} to \codamosa{} in Section~\ref{evaluation:comparison}.

\subsubsection*{LLM-based Approaches}
Barei\ss{} et al.\ study the performance of the Codex LLM on Java test case generation, among other code generation tasks~\cite{bareiß2022code}.
Its prompts contain the signature of the method under test, an example of test generation, and the method's body; it discards any tests that do not compile.
By contrast, \coverup{}'s prompts are based on code segments lacking coverage, and they explicitly request tests to improve it.
\coverup{} also continues the chat with the LLM in case of build failure, failing tests, or lack of coverage.
Section~\ref{evaluation:continued-dialogue} shows that this iterative dialogue is responsible for nearly 40\% of \coverup{}'s successful test generation.

Vikram et al.\ discuss prompting LLMs based on API documentation to generate property-based tests for Python~\cite{vikram-pbt}. 
\coverup{} instead bases its prompting on the source code and coverage measurements, and while it accepts property-based tests if the LLM generates them, it does not explicitly request them.

\textsc{TiCoder} prompts LLMs to generate tests based on a natural language description of the intended functionality of code~\cite{ticoder}.
Rather than facilitate regression testing, however, \textsc{TiCoder} generates tests to clarify and formalize user intent.

\textsc{TestPilot} prompts an LLM to generate JavaScript unit tests based on the function under test's implementation, its documentation, and usage snippets~\cite{testpilot}.
Like \coverup{}, \textsc{TestPilot} checks the tests generated by the LLM and continues the chat to refine the prompt in case of errors, but it does not prompt based on coverage, nor does it continue the chat if the new tests do not improve coverage, nor does it use tool functions to provide the LLM with additional context.

ChatUniTest~\cite{chatunitest} prompts LLMs to generate Java unit tests.
Its use of LLMs is limited to prompting with code and for repairs when a generated test fails compilation.
Unlike \coverup{}, ChatUniTest does not employ coverage measurements to indicate what lines or branches lack coverage, nor does it request improvements if the generated tests do not improve coverage.

\subsubsection*{Concurrent Work Using LLM-based Approaches}
The approaches below were developed concurrently with \coverup{}, which was initially posted on GitHub on August 7, 2023:

\textsc{Fuzz4All}~\cite{fuzz4all} uses two separate LLMs to \emph{fuzz test} programs in various programming
languages.
The \emph{distillation} LLM takes in arbitrary user input, such as documentation and code examples, and generates prompts for the \emph{generation} LLM, which produces test inputs.
After the initial user input distillation and prompt selection, \textsc{Fuzz4All} repeatedly employs the generation LLM, mutating its prompt to produce additional test inputs.
While \textsc{Fuzz4All} and \coverup{} both use LLMs to generate test cases, their approaches differ fundamentally: \textsc{Fuzz4All} generates test inputs based on documentation or examples, relying on a user-provided test oracle for testing; \coverup{} instead works based on the source code and coverage measurements, looking to add tests that improve the test suite's coverage.

SymPrompt~\cite{symprompt} focuses on generating tests that cover hard-to-reach code.
To do so, it first determines the path constraints needed to reach a certain part of the code and then prompts the LLM for tests based on those constraints.
CoverUp's tests can also cover such code, but by prompting based on coverage measurements, a simpler and more direct approach.
While SymPrompt extracts related declarations and includes them statically in its prompts, \coverup{} combines statically generated \verb|import| declarations with a tool function that enables the LLM to drive the code context discovery process.  
SymPrompt's error handling is limited to deleting lines in the response in an attempt to correct syntax errors; \coverup{} continues the chat in case of errors, asking the LLM for a correction.
\coverup{} also continues the chat if the test does not improve coverage.
SymPrompt is evaluated on an undisclosed subset of 897 functions (from at least 4,546) previously used by \codamosa{} and BugsInPy~\cite{Widyasari_2020}.
It achieves 74\% line coverage using OpenAI's GPT-4 LLM, approximately a 2x improvement over its baseline prompt~\cite{symprompt}.
Without knowing the exact functions employed despite multiple requests for code and/or a replication package\footnote{Personal communication with SymPrompt authors.}, we can only provide a coarse comparison: as Section~\ref{evaluation:comparison} shows, using GPT-4o \coverup{} obtains 82\% median per-module line coverage and 64\% overall line coverage on a 4,116-function suite also derived from \codamosa{}'s evaluation.
Even though \coverup{} is evaluated on a 5x larger suite, the median per-module coverage it achieves is also approximately a 2x improvement over an ablated version of itself (Section~\ref{evaluation:comparison}).

TestGen-LLM~\cite{meta-unit-tests} is a test generation tool developed and deployed at Meta to improve Kotlin unit tests.
TestGen-LLM prompts LLMs for tests that improve coverage, including in the prompt the class under test and, in some cases, the existing test class.
It discards any tests that do not build, do not pass, or do not improve coverage and presents any remaining tests to a developer for approval during code review.
Like TestGen-LLM, \coverup{} rejects tests that fail or lack coverage, but instead of simply discarding them, it continues the chat with the LLM, asking for improvements.
\coverup{} also indicates in its prompts what portions of the code lack coverage and allows the LLM to discover additional context through a tool function.

MuTAP~\cite{mutap} prompts an LLM for Python unit tests based on mutation testing.
MuTAP first prompts for a test for a portion of code and, using it, performs mutation testing.
It then prompts for new assertions for each surviving mutant, adding these to the test.
Like \coverup{}, MuTAP also prompts the LLM for a repair in case of syntax errors, but unlike \coverup{}, it does not prompt it for improvements in case of other errors or lack of coverage, nor does it include coverage information in its prompts.
Unfortunately, MuTAP currently lacks any ability to provide context for the code under test and hardcodes the datasets used for evaluation in the code, hampering efforts to evaluate it on other code.
In contrast, our implementation of \coverup{}, like Pynguin and \codamosa{}, can be applied to any Python package.
We empirically compare \coverup{} to MuTAP in Section~\ref{evaluation:comparison}.
\begin{figure*}[t]
    \includegraphics[width=\textwidth]{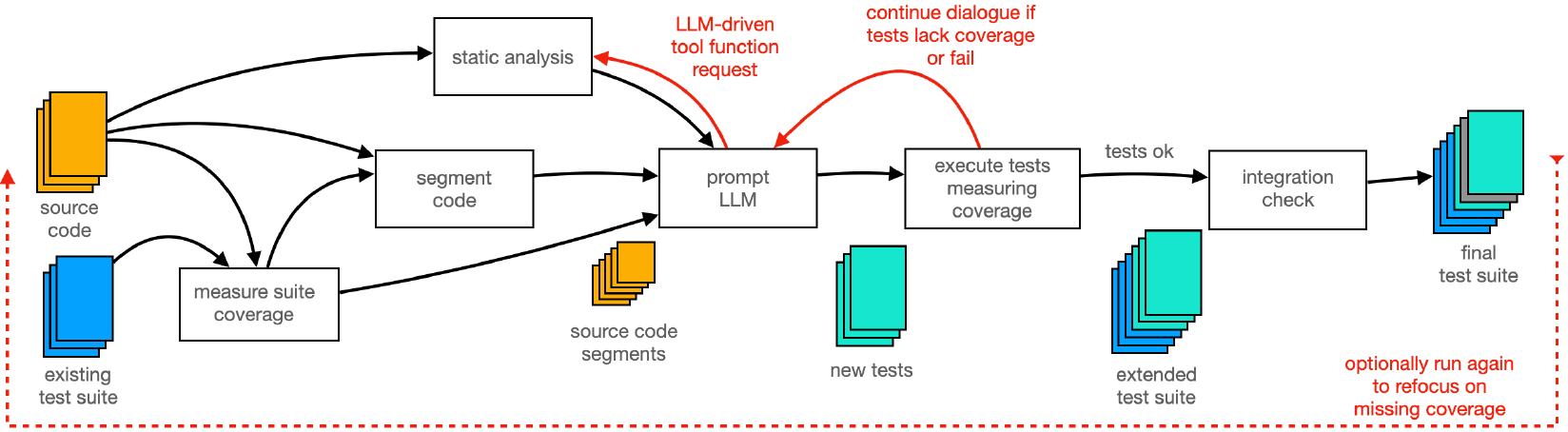}
    \vspace{0.5em}
    \caption{
        \textbf{Graphical Overview of the \coverup{} Algorithm.}
        \textnormal{
            After measuring the test suite's coverage, \coverup{} prompts the LLM, focusing it on code segments that lack coverage.
            It checks each new test, either accepting it if it increases coverage or continuing the dialogue if it needs improvements.
            \coverup{} provides a tool function with which the LLM can request additional context.
            A final check helps ensure that the suite works as a whole.
            Given its incremental nature, once done, \coverup{} can be rerun to refocus the LLM on any segments missed in previous passes.
        }
        \label{fig:overview}
    }
\end{figure*}

\section{Approach}
This section presents a detailed description of \coverup{}.
Figure~\ref{fig:overview} provides a graphical overview of its algorithm.
\coverup{} first measures the code coverage obtained by the existing test suite and uses it to identify code segments (functions and/or classes) that require additional testing (Section~\ref{impl:code-segmentation}).
It then prompts an LLM for tests for each segment, combining information from coverage analysis and static code analysis (Section~\ref{impl:prompting}).
\coverup{} next executes the LLM-generated tests, once again measuring coverage.
If the tests do not compile, fail to run, or do not increase coverage, \coverup{} continues the chat with prompts that request improvements and include any error messages (Section~\ref{impl:dialogue}).
In both initial and continued prompts, \coverup{} offers the LLM a tool function with which it can request additional contextual information about names in the source code, such as function or type definitions (Section~\ref{impl:tool-function}).
Having handled all code segments, \coverup{} checks the extended test suite, looking for any integration problems (Section~\ref{impl:integration-check}).
The entire process can be repeated arbitrarily, enabling the LLM to refocus on segments not covered in previous iterations.

\subsection{Code Segmentation}\label{impl:code-segmentation}
\coverup{}'s first step is to measure the code coverage of any pre-existing test suite.
It then uses the AST of each source file missing coverage to identify code segments that need additional coverage.
Code segments typically consist of a single function or method; subject to a size limit, however, they may contain an entire class.
As Figure~\ref{fig:class-excerpt} shows, if a segment contains a method, \coverup{} also includes a few lines of the original class definition to provide the LLM with context but omits other methods and definitions.
We find that retaining comments improves code comprehension; for that reason, \coverup{} does not remove them, thus trading a loss in concision for an increase in generation quality.

The goal is to provide the LLM, with each prompt, a code excerpt that is intelligible, provides enough context, includes the lines or branches lacking coverage, and is as short as possible.
Keeping code segments short is important primarily because of LLMs' limits on their context (input) window.
While various current LLMs, such as OpenAI's GPT-4 and GPT-4o, Meta's LLaMA 3, and Anthropic's Claude 3, support large context windows of over 100K tokens, this limit applies to more than just the initial prompt.
It applies to the number of tokens in the entire sequence of messages included with each request, which grows with each tool function call, LLM response, and continued chat prompt.
Even when prompts fit in these long context windows, shorter prompts remain preferable: Rosas et al.~\cite{rosas2024aioptimizecodecomparative} show that as the size of prompts containing code increases, so does the likelihood of inaccuracies.
Finally, LLM providers typically charge on a per-token basis, so it literally pays to be succinct.

Concretely, to identify the code segments in a source file, \coverup{} first computes a set of ``interesting'' lines: these are lines that either lack coverage or are the source or the destination in branches that lack coverage.
It then looks in the AST for a class, function, or method object containing each line.
If the object found is a class that spans more lines than a configurable limit, \coverup{} adds that class definition as segment context and recursively looks for another, smaller object containing the line.
Algorithm~\ref{alg:code-segments} describes this process in more detail.

We implement this step using SlipCover~\cite{slipcover}, a recently introduced coverage analyzer with near-zero overhead, and Python's \texttt{ast} module.

\begin{algorithm}
\DontPrintSemicolon 
\SetFuncSty{textsc}
\SetAlCapSkip{1cm}

\SetKwFunction{FIdentifySegments}{identify\_segments}
\SetKwFunction{FMissingLines}{missing\_lines}
\SetKwFunction{FLinesIn}{lines\_in}
\SetKwFunction{FMissingBranches}{missing\_branches}
\SetKwFunction{FParseAst}{parse\_ast}
\SetKwFunction{FFindLine}{find\_line}
\SetKwProg{Fn}{Function}{}{}

\Fn{\FIdentifySegments{coverage, max\_len}}{
    $code\_segs \gets \emptyset$\;
    \ForEach{$file$ \textbf{in} $coverage.files$}{
        $interesting \gets \FMissingLines{coverage, file}~\cup$
        $\FLinesIn{\FMissingBranches{coverage, file}}$\;
        $ast \gets \FParseAst{file}$\;
        \ForEach{$line$ \textbf{in} $interesting$}{
            $context \gets$ empty list\;
            $node \gets \FFindLine{ast, line}$\;
            \While{($node$ \textbf{is-a} Class) \textbf{and}
                $\textsc{length}(node) > max\_len$ \textbf{and}
                $(inner \gets \FFindLine{node, line})$}{
                append $node$ to $context$\;
                $node \gets inner$\;
            }
            $code\_segs \gets code\_segs \cup \{node, context\}$\;
        }
    }
    \Return $code\_segs$\;
}
\caption{Algorithm for identifying the code segments lacking coverage, based on a coverage measurement and a tentative maximum segment length.}
\label{alg:code-segments}
\end{algorithm}

\begin{figure}
\begin{lstlisting}[language=Python,numbers=left,xleftmargin=10pt]
class AnsiTextWrapper(TextWrapper):
  def _wrap_chunks(self, chunks: List[str]) -> List[str]:

    lines = []
    if self.width <= 0:
      raise ValueError("invalid width %r (must be > 0)" % self.width)
    if self.max_lines is not None:
      if self.max_lines > 1:
        indent = self.subsequent_indent
    [...]
\end{lstlisting}
    \caption{\textbf{\coverup{} summarizes method excerpts:}
        \textnormal{
            \coverup{} generates compact code excerpts.
            In this case, 200 lines of source code were originally present between listing lines 1 and 2.
            The original code is from the \texttt{flutils} package.
        }
        \label{fig:class-excerpt}
    }
\end{figure}

\subsection{Initial Prompting}\label{impl:prompting}
\coverup{} next prompts the LLM for tests for each code segment it identified in the previous step.
Figure~\ref{fig:initial-prompt} shows an example of an initial prompt, with circled numbers identifying sections.
It has the following structure:
\begin{itemize}
    \item[\circled{orange!15}{1}] a statement assigning the LLM the \emph{persona}~\cite{prompt-catalog} of an ``expert Python test-driven developer'', intended to help guide it towards high quality tests;

    \item[\circled{green!15}{2}] a sentence pointing out the code excerpt (segment), identifying what file it comes from, and stating what lines or branches do not execute when tested.
        The portion specifying the lines and branches missing coverage is compressed using line ranges, simplifying the prompt and reducing token usage.

    \item[\circled{red!15}{3}] a request for \texttt{pytest} test functions and an encouragement for the LLM to use the provided tool function;

    \item[\circled{blue!15}{4}] a series of other requests, such as ``include assertions'' and ``avoid state pollution'', to steer the result towards usable tests;

    \item[\circled{gray!15}{5}] a request that the response only include the new Python tests to facilitate its extraction from the response and to reduce token usage; and

    \item[\circled{yellow!15}{6}] the code segment, prefixed by generated \texttt{import} statements
        and tagging the lines lacking (line or branch) coverage with their numbers.
\end{itemize}

\begin{figure}
    \begin{colorlisting}{orange!15}{1}{}
You are an expert Python test-driven developer.
    \end{colorlisting}
    \begin{colorlisting}{green!15}{2}{}
The code below, extracted from thonny/plugins/pgzero_frontend.py, does not achieve full
coverage: when tested, line 19 does not execute.
    \end{colorlisting}
    \begin{colorlisting}{red!15}{3}{}
Create new pytest test functions that execute all missing lines and branches, always making
sure that each test is correct and indeed improves coverage.
Use the get_info tool function as necessary.
    \end{colorlisting}
    \begin{colorlisting}{blue!15}{4}{}
Always send entire Python test scripts when proposing a new test or correcting one you
previously proposed.
Be sure to include assertions in the test that verify any applicable postconditions.
Please also make VERY SURE to clean up after the test, so as to avoid state pollution;
use 'monkeypatch' or 'pytest-mock' if appropriate.
Write as little top-level code as possible, and in particular do not include any top-level
code calling into pytest.main or the test itself.
    \end{colorlisting}
    \begin{colorlisting}{gray!15}{5}{}
Respond ONLY with the Python code enclosed in backticks, without any explanation.
    \end{colorlisting}
    \begin{colorlisting}{yellow!15}{6}{}
```python
            import os
            from thonny import get_workbench
            def update_environment():
                if get_workbench().in_simple_mode():
                    os.environ["PGZERO_MODE"] = "auto"
                else:
        19:         os.environ["PGZERO_MODE"] = str(get_workbench().get_option(_OPTION_NAME))
```

    \end{colorlisting}
    \caption{\textbf{Example of an initial prompt:}
        \textnormal{
        The initial prompt selects a persona (1),
        identifies where the code comes from and states how it lacks coverage when tested (2),
        asks for tests (3-5),
        and shows the code segment, prefixed by \texttt{import} statements and
        with the line lacking coverage identified by its number.
        The code shown is from the \texttt{thonny} package.
        }
        \label{fig:initial-prompt}
    }
\end{figure}

We find that tests generated by GPT-4 often include top-level code calling into \texttt{pytest.main} or into parts of the test itself.
While such top-level code can make sense in a standalone test file, Python executes it as part of the loading process, and doing so may significantly disrupt \texttt{pytest}'s operation.
In fact, in some of these early test generations, such calls caused \texttt{pytest} to restart its test discovery, slowing it down until it became unusable.
For that reason, part \circled{blue!15}{4} of the prompt directs the LLM not to include such calls.

Prompt part \circled{blue!15}{4} also instructs the LLM to generate tests that ``clean up [...], so as to avoid state pollution''.
It suggests using \texttt{monkeypatch} and \texttt{pytest-mock}, useful for automatically cleaning up after tests and isolating the software under test from other parts of the code.
Including these instructions improves the generated tests; nonetheless, we still observe test generation that includes side effects.
Section~\ref{impl:integration-check} describes how \coverup{} handles such tests.

Prior to the code excerpt in part \circled{yellow!15}{6}, \coverup{} adds \texttt{import} statements that provide context for each code segment: we find that without these, LLMs often make incorrect assumptions about symbols in the code, leading to errors.
Rather than extract these verbatim from source code, \coverup{} generates imports based on static analysis.
In particular, it converts any relative imports to absolute imports.
Using relative imports in code excerpts can lead to tests that use them outside of package scope, where they are invalid.

Within the code excerpt, \coverup{} tags certain lines with their numbers, as in ``\verb|19:|'' in Figure~\ref{fig:initial-prompt}.
It tags all lines lacking coverage and those that are part of a branch lacking coverage.
This tagging improves the LLM's understanding of the missing coverage: we find that prompts that only indicate the starting line number at the beginning of the excerpt lead to tests that do not improve coverage.

To send the prompt using OpenAI's API, \coverup{} embeds it as a ``message'' in a JSON-formatted request that also includes other fields specifying the model to use, meta-parameters such as the model temperature, etc.
Figure~\ref{fig:json-request} shows an example.
\begin{figure}
    \begin{lstlisting}[language=Python,xleftmargin=10pt]
{
  "model": "gpt-4o-2024-05-13",
  "temperature": 0,
  "messages": [
    {
      "role": "user",
      "content": """
You are an expert Python test-driven developer.
The code below, extracted from code/funcs.py, does not achieve full coverage:
when tested, line 4 does not execute.
[...]
```python
      from code import A
      def func(a: A) -> int:
4:        return bool(a.x > 5 or a.x < 2)
```
"""
    }
  ]
}
    \end{lstlisting}
    \caption{\textbf{Sending a prompt:} to prompt and LLM using OpenAI's chat API, \coverup{} puts together a JSON-formatted request that includes the prompt in ``messages''.
        \texttt{[...]} indicates a portion omitted for brevity.
    }
    \label{fig:json-request}
\end{figure}

\subsection{Verification and Continued Chat}\label{impl:dialogue}
Once the LLM generates tests in response to the initial prompt, \coverup{} executes them, again measuring coverage.
In our implementation, this step is made more efficient by SlipCover's near-zero overhead: using \verb$coverage.py$, the only other alternative tool for Python, introduces up to 260\% overhead~\cite{slipcover}.

If the new tests pass and increase coverage, \coverup{} saves them.
If, conversely, they do not increase coverage or result in failures or errors, \coverup{} continues the chat session, pointing out the problem(s) and requesting improvements.
To continue the chat, \coverup{} sends another request to the LLM, appending the LLM's response and a new prompt to the previous messages.
Figure~\ref{fig:continued-chat} shows how \coverup{} continues the chat; Figures~\ref{fig:follow-up-coverage} and \ref{fig:follow-up-error} show examples of prompts requesting improvements.

\begin{figure}
    \begin{lstlisting}[language=Python,xleftmargin=10pt]
{
  "model": "gpt-4o-2024-05-13",
  "temperature": 0,
  "messages": [
    {
      "role": "user",
      "content": """
You are an expert Python test-driven developer.
The code below, extracted from code/funcs.py, does not achieve full coverage:
[...]
"""
    },
    {
      "role": "assistant",
      "content": """
```python
import pytest

def test_something():
    [...]
"""
    },        
    {
      "role": "user",
      "content": """
Executing the test yields an error, shown below.
Modify the test to correct it; respond only with the complete Python code in backticks.
[...]
"""
    }
  ]
}
    \end{lstlisting}
    \caption{\textbf{Continuing the chat:} to continue a chat, \coverup{} sends out a new request, including the previous messages (here, the initial prompt), the LLM's response (indicated with the ``assistant'' role), and the new prompt.
        \texttt{[...]} indicates a portion omitted for brevity.
        \label{fig:continued-chat}
    }
\end{figure}

\begin{figure}
    \begin{colorlisting}{green!15}{1}{}
This test still lacks coverage: line 615 and branches 603->exit, 610->608, 618->exit
do not execute.
    \end{colorlisting}
    \begin{colorlisting}{red!15}{2}{}
Modify it to correct that; respond only with the complete Python code in backticks.
Use the get_info function as necessary.
    \end{colorlisting}
    \caption{\textbf{Example of a coverage follow-up prompt:}
        \textnormal{
        \coverup{} indicates to the LLM that a line and
        some branches still weren't covered (1), asking that it correct the test (2).
        }
        \label{fig:follow-up-coverage}
    }
\end{figure}

\begin{figure}
    \begin{colorlisting}{green!15}{1}{}
Executing the test yields an error, shown below.
    \end{colorlisting}
    \begin{colorlisting}{red!15}{2}{}
Modify the test to correct it; respond only with the complete Python code in backticks.
Use the get_info function as necessary.
    \end{colorlisting}
    \begin{colorlisting}{yellow!15}{3}{breaklines=true}
    def test_AnsiTextWrapper_non_stripped_placeholder():
        wrapper = AnsiTextWrapper()
        wrapper.placeholder = " \x1b[31mplaceholder\x1b[0m "
>       assert wrapper.placeholder_len == len_without_ansi(wrapper.placeholder.strip())
E       AssertionError: assert 13 == 11
E        +  where 13 = <flutils.txtutils.AnsiTextWrapper object at 0x7f992ca5e440>.placeholder_len
E        +  and   11 = len_without_ansi('\x1b[31mplaceholder\x1b[0m')
E        +    where '\x1b[31mplaceholder\x1b[0m' = <built-in method strip of str object at 0x7f992bc06c90>()
E        +      where <built-in method strip of str object at 0x7f992bc06c90> = ' \x1b[31mplaceholder\x1b[0m '.strip
E        +        where ' \x1b[31mplaceholder\x1b[0m ' = <flutils.txtutils.AnsiTextWrapper object at 0x7f992ca5e440>.placeholder

coverup-tests/tmp_test_s4dcdnes.py:7: AssertionError
    \end{colorlisting}
    \caption{\textbf{Example of an error follow-up prompt:}
        \textnormal{
        \coverup{} indicates to the LLM that an error occurred (1), requests a repair (2), and includes an excerpt of the error messages (3).
        In the original execution from which this example is taken, the LLM responds with a usable test (not shown).
        }
        \label{fig:follow-up-error}
    }
\end{figure}

Before executing a new set of tests, \coverup{} looks for any Python modules used by the test that are absent from the system.
These missing modules are typically test helper modules, such as the \texttt{pytest-ansible} plugin used to help test the \texttt{ansible} package.
Our implementation offers options to install missing modules automatically and record these in Python's \texttt{requirements.txt} format to facilitate their use in a subsequent run.

\subsection{Tool Functions}\label{impl:tool-function}
Tool functions allow an LLM to interact with external tools~\cite{openai-function-calling}.
\coverup{} exposes a \texttt{get\_info} tool function, which allows the LLM to request additional information about any names in the excerpt, such as types or variables.
In response, \coverup{} provides a portion of the source code that shows the definition of the requested object.

To indicate support for the \texttt{get\_info} tool function, \coverup{} includes its description in a ``tools'' JSON element within each chat request.  
If the LLM needs to invoke this function, it may respond with a list of \texttt{get\_info} call requests instead of generating text, each specifying the symbol to retrieve.  
\coverup{} then continues the chat by sending a new request that appends the LLM’s call requests and the retrieved results from \texttt{get\_info} to the previous messages.

\begin{figure}
    \begin{lstlisting}[language=Python,xleftmargin=10pt]
You are an expert Python test-driven developer.
[...]
Use the get_info tool function as necessary.
[...]
```python
            from code import A
            def func(a: A) -> int:
         4:     return bool(a.x > 5 or a.x < 2)
```
    \end{lstlisting}
    \caption{\textbf{Example of a prompt likely to lead to a tool function call:}
        \textnormal{
        In response to this initial prompt, the LLM would likely choose to respond with a request to call \texttt{get\_info("code.A")}, as it requires more information on how to instantiate \texttt{A} in tests.
        \texttt{[...]} indicates a portion omitted for brevity.
        }
        \label{fig:prompt-for-function-call}
    }
\end{figure}

\begin{figure}
    \begin{lstlisting}[language=Python,xleftmargin=10pt]
"..." below indicates omitted code.

```python
class A:
    def __init__(self, *, initial_x=0):
        self._x = initial_x

    @property
    def x(self):
        ...
```
    \end{lstlisting}
    \caption{\textbf{Example function call response:}
        \textnormal{
        Asked about \texttt{code.A}, \coverup{} performs static analysis and, discovering that \texttt{A} is a class, responds with an abbreviated form of its definition. 
        In this case, it includes its \texttt{\_\_init\_\_} constructor, which indicates to the LLM that \texttt{A}'s constructor takes the \texttt{initial\_x} keyword argument.
        The response does not include any of the contents of method \texttt{x}; the LLM could subsequently ask for \texttt{code.A.x} to obtain more information on it.
        The \texttt{...} in the Python excerpt is literal: it is a valid no-op token in Python, and \coverup{} uses it to indicate where code has been removed.
        }
        \label{fig:function-call-response}
    }
\end{figure}

For example, when given the prompt shown in Figure~\ref{fig:prompt-for-function-call}, the LLM might ask for information on \verb$code.A$.
The function response, shown in Figure~\ref{fig:function-call-response}, indicates that values of \verb$x$ must be passed to the class constructor using a keyword argument.
This response lets the LLM immediately respond with correct tests; we find that if we do not include the tool function in this situation, the LLM attempts to pass values of \verb$x$ to the constructor, leading to errors.

To generate its response, \verb$get_info$ first looks for a function, class, module, or variable matching the requested name.
Supporting variables is particularly important, as constants and aliases in Python are created using variable assignments.
Beginning with the module where the excerpt originated, \verb$get_info$ may extend its search to other modules by following \texttt{import} statements.
Once it locates the definition, \verb$get_info$ constructs a response by extracting relevant portions from each module.  
For many constructs, such as assignments, \texttt{import} statements, or function definitions, it includes the entire definition. 
However, for classes and modules, including their full definitions may result in prohibitively long responses.  
Simply omitting elements, as done in the initial prompt, can lead to missing essential information or cause the LLM to assume such elements do not exist.  
For example, if \verb$get_info$ had entirely omitted property \verb|x| in Figure~\ref{fig:function-call-response}, there would be no indication of how attribute \verb|A.x| was accessed.
Instead, \verb|get_info| replaces the bodies of such elements with \verb|...|, a valid ``no-op'' token in Python, and indicates the omission in the response.

\coverup{} makes \verb$get_info$ available both with initial (Section~\ref{impl:prompting}) and continuation prompts (Section~\ref{impl:dialogue}) so that it can help generate both better responses and better corrections.

\subsection{Integration Check}\label{impl:integration-check}
While \coverup{} only saves tests that pass when run in isolation, it is still possible that some may fail to clean up after themselves or introduce side effects (``state pollution'') that cause other tests to fail~\cite{10.1145/2771783.2771793}.
This issue is not unique to LLM-based test generation: Fraser et al.~\cite{evosuite-evaluation} report side effects on 50\% of test classes generated by \textsc{EvoSuite}~\cite{evosuite}, an SBST test generator.
Figure~\ref{fig:test-with-side-effects} shows an example where a generated test not only modifies but also deletes global symbols used by the code under test, in a misguided attempt to clean up.
For a test generation approach to be effective, it must thus include some way to handle state pollution.

Our implementation offers three alternatives:
\begin{enumerate}
    \item \textbf{Execute each test in isolation:}
        This approach prevents memory-based state pollution by one test from affecting other tests.
        \coverup{} implements efficient test isolation based on the Unix \verb$fork()$ system call for the \verb$pytest$ framework.
        This option is the default, which Section~\ref{sec:evaluation} uses for the empirical evaluation of \coverup{}.
    \item \textbf{Disable polluting tests:}
        This approach executes the entire suite and, upon a test failure, searches for the polluting tests and disables them.
        \coverup{} implements the search by successively reducing and executing subsets of the entire suite;
    \item \textbf{Disable failing tests:}
        This approach may lead to a significant loss in coverage as tests may be disabled unnecessarily and may still leave a polluting test enabled, but it allows the user to move on quickly to prompting for more tests.
\end{enumerate}
Any disabled tests remain available for review and possible reactivation by the user.

\begin{figure}
\begin{lstlisting}[language=Python,numbers=left,xleftmargin=10pt]
from ansible import constants as C

def test_process_include_results():
    C._ACTION_ALL_INCLUDES = ['include', 'include_tasks', 'import_tasks', 'import_playbook']
    C._ACTION_INCLUDE = 'include'

    assert [...]

    del C._ACTION_ALL_INCLUDES
    del C._ACTION_INCLUDE
\end{lstlisting}
    \caption{\textbf{Handling state pollution:}
        \textnormal{
            This LLM-generated test overwrites global constants and then deletes them in a misguided attempt to clean up.
            The test succeeds if executed by itself, but when executed along with other tests, the missing constants cause other tests to fail.
            Our implementation offers three alternatives to handle this situation: it can execute tests in isolation (the default), disable the polluting tests, or disable the failing tests.
            \texttt{[...]} indicates a portion omitted for brevity.
        }
        \label{fig:test-with-side-effects}
    }
\end{figure}

\subsection{Handling Flaky Tests}\label{impl:handling-flakies}
Flaky tests are those that can both pass or fail inconsistently without any changes to the code under test~\cite{flaky-tests-survey, 10.1145/3428270, 10.1145/3597503.3608138}.
Their unreliable behavior has a variety of causes; in the context of LLM-generated tests, they can also, in part, result from not providing the LLM with sufficient information.
Figure~\ref{fig:flaky-random-test} shows a test for the \texttt{mimesis} data generator package.
The code under test randomly assigns a blood type to its \texttt{Person} object; the LLM, not having been provided with the list of valid names, assumes that it includes \texttt{0-} and \texttt{0+} whereas code actually uses \texttt{O-} and \texttt{O+} (using the letter O, not zero).
Consequently, the test fails whenever that blood type is assigned.

\begin{figure}
\begin{lstlisting}[language=Python,numbers=left,xleftmargin=10pt]
from mimesis.providers.person import Person

def test_blood_type():
    blood_type = Person().blood_type()    # this performs a random assignment
    [...]
    BLOOD_GROUPS = ['0-', '0+', 'A-', 'A+', 'B-', 'B+', 'AB-', 'AB+']
    assert blood_type in BLOOD_GROUPS
\end{lstlisting}
    \caption{\textbf{Example of a flaky test:}
        \textnormal{
            This test checks that a randomly assigned blood type is valid.
            It fails whenever the blood type is \texttt{O+} or \texttt{O-}, as the LLM assumed these would be named using zero rather than the letter O.
            \coverup{} executes each test a few times, making the error more likely to surface and thus allowing the LLM to attempt to correct it.
        }
        \label{fig:flaky-random-test}
    }
\end{figure}

Our implementation uses the \verb|pytest-repeat| module to execute each newly generated test multiple times, making any flaky tests more likely to fail.
If a test fails, \coverup{} continues the chat, allowing the LLM to attempt to correct the problem.

\subsection{Other Technical Challenges}\label{impl:technical-challenges}
Making \coverup{} a practical tool poses several technical challenges.
One such challenge stems from the time spent in LLM inference and executing tests.
Each individual prompt sent through OpenAI's API typically requires several seconds to complete.
Similarly, executing individual LLM-generated tests and measuring the new coverage achieved requires additional time.
\coverup{} repeats this process for each code segment and each time the dialogue is continued.
Even though our implementation limits the time spent waiting on responses and test executions, if each code segment were processed serially, creating tests for packages of nontrivial size would take an unacceptable amount of time.
Instead, it prompts for tests and verifies them asynchronously, using the Python \texttt{asyncio} package.
As a result, during \coverup{}'s evaluation (Section~\ref{sec:evaluation}), we observe a 500x speedup over serial execution.

Other practical challenges arise because OpenAI's API is provided as a cloud service and is subject to various limits.
Our implementation handles various timeout and other error conditions automatically.
To avoid exceeding the rate limits imposed by OpenAI, it spreads out its requests using a leaky bucket scheme~\cite{leaky-bucket} implemented by the Python module \verb|aiolimiter|.
Additionally, we implement checkpointing to files, allowing the user to resume prompting for tests (and not lose any progress so far) after interrupting it or stopping due to unforeseen circumstances, such as the OpenAI account running out of funds.

    \section{Evaluation}
\label{sec:evaluation}

Our evaluation investigates the following questions:
\begin{description}
    \item[\textbf{RQ1:}] Does the coverage of \coverup{}'s generated tests improve upon the state of the art?
    (Section~\ref{evaluation:comparison})
    \item[\textbf{RQ2:}] How effective is \coverup{} compared to simply prompting an LLM for tests?
    (Section~\ref{evaluation:just-prompt})
    \item[\textbf{RQ3:}] How effective are \coverup{}'s continued dialogues at increasing coverage?
    (Section~\ref{evaluation:continued-dialogue})
    \item[\textbf{RQ4:}] How does the cost of running \coverup{} compare to \codamosa{}?
    (Section~\ref{evaluation:cost})
    \item[\textbf{RQ5:}] How important are \coverup{}'s components to its performance?
    (Section~\ref{evaluation:component-ablations})
\end{description}

\subsection{Experimental Setup}\label{evaluation:setup}

\subsubsection*{\textbf{Benchmarks}}
We utilize three benchmark suites:
\begin{itemize}
    \item \emph{CM}, a benchmark suite on which Pynguin struggles to obtain high coverage.
        Collated originally by the authors of \codamosa{}~\cite{codamosa} and available from \url{https://github.com/microsoft/codamosa}, it is derived from 35 open-source projects used in the evaluation of BugsInPy~\cite{Widyasari_2020} and Pynguin~\cite{pynguin-empirical-study}.
        It contains $\approx$100,000 lines of code across 425 Python modules.

    \item \emph{PY}, a set of modules originally excluded from \codamosa{}'s suite because Pynguin already performs well on it.
        We evaluate \coverup{} on these modules so as not to leave open the question of how well it performs on code where Pynguin / SBST already performed well.
        It contains $\approx$5,000 lines of code across 84 Python modules.

    \item \emph{MT}, a dataset of functions extracted from the HumanEval dataset~\cite{humaneval}, originally used in the evaluation of MuTAP~\cite{mutap}.
        We use this dataset to enable comparisons to MuTAP since its implementation lacks the support needed to run on arbitrary Python packages.
        It contains $\approx$2,000 lines of code across 163 functions.

\end{itemize}

CM excludes the \texttt{mimesis}, \texttt{sanic}, and \texttt{thef*ck} packages from the original \codamosa{} suite: \texttt{mimesis} is a package for generating random data, making it challenging to generate non-flaky tests for it.
\texttt{mimesis} was included in \codamosa{}'s original evaluation, but the tests \codamosa{} generates do not contain any assertions and thus do not fail in the face of randomly generated values.
The \texttt{sanic} and \texttt{thef*ck} packages require modules that either are no longer available or conflict with \coverup{}.

\subsubsection*{\textbf{Baselines}}
To examine RQ1, we compare against two versions of \codamosa{} and four versions of MuTAP, varying their LLM and prompt type:
\begin{itemize}
    \item \emph{\codamosa~(codex)}, the original version that uses the Codex LLM.
        Since Codex is no longer available, we use original tests from \codamosa's evaluation, using its best performing configuration (\verb|0.8-uninterp|); 

    \item \emph{\codamosa~(gpt4o)}, our adapted version that uses the same model as \coverup{}, to help rule out differences in performance due to the use of different LLMs.

    \item \emph{MuTAP~(codex few-shot)}, the original version that uses the Codex LLM, using a ``few-shot'' prompt, where two examples of test generation are included with the prompt.
        Since Codex is no longer available, we use original tests from MuTAP's evaluation; 

    \item \emph{MuTAP~(codex zero-shot)}, also the original Codex version, but using
        a prompt without examples.
        Here, too, we use original tests from MuTAP's evaluation;

    \item \emph{MuTAP~(gpt4o few-shot)}, our adapted version that uses the same model as \coverup{}, with newly generated tests using a ``few-shot'' prompt;

    \item \emph{MuTAP~(gpt4o zero-shot)}, our adapted version, but using a ``zero-shot'' prompt.
\end{itemize}

To create the gpt4o versions of \codamosa{} and MuTAP, we modify these to use OpenAI's chat API and, in the case of \codamosa{}, insert instructions requesting a code completion before its original code completion prompt.
For MuTAP, we insert a prompt requesting that it respond in the same format as a Codex model.

We could not empirically evaluate~\cite{symprompt} as, at the time of writing, it remains unavailable: its authors have indicated that it is not publicly available.

For RQ2, we create an ``\coverup~(ablated)'', a version of \coverup{} which utilizes a nearly identical prompt, except in that it does not specify how the code lacks coverage or tags lines lacking coverage, does not add computed \texttt{import} statements, does not offer a tool function for additional context, and does not continue the chat in case of errors or lack of coverage.

\subsubsection*{\textbf{Metric}}\label{eval:metric}
We utilize the line, branch, and combined line and branch coverage as metrics, computing these for all benchmark modules.
We also compute the combined line and branch coverage on a per-module basis.
It is necessary to include both line and branch coverage as metrics because in Python, branch coverage does not subsume line coverage: various situations can lead the Python interpreter to throw exceptions, and when thrown, these exceptions are not recorded as branches in coverage information.
We run all generated tests in isolation (see Section~\ref{impl:integration-check}) and measure all coverage using SlipCover~\cite{slipcover}.

\subsubsection*{\textbf{Execution Environment}}\label{eval:execution-environment}

We evaluate both \codamosa{} and \coverup{} using the \texttt{codamosa-docker} Docker image available at \url{https://github.com/microsoft/codamosa}, modified only to install SlipCover and to disable its default \texttt{entrypoint} script, allowing easy execution of other scripts.
The image is based on Debian 11 and includes Python 3.10.2, with which we run all benchmarks.
As the host system for Docker, we utilize a Linux kernel 5.6 system with 10 Intel i9 cores at 3.7GHz and 64GB RAM.

Before each measurement, our scripts install the benchmark-specific \texttt{package.txt} requirements file distributed with \codamosa{} using \texttt{pip}.
Unfortunately, \texttt{pip} fails to install some of these requirements.
Since these failures were originally ignored for \codamosa{}, we also ignore them to replicate the original conditions as closely as possible.

For MuTAP and the tests it generates, we use Python 3.9.12: MuTAP uses MutPy~\cite{mutpy} for mutation testing, which requires older versions of Python.

\subsubsection*{\textbf{CoverUp options}}
We configure \coverup{} to use OpenAI's \verb|gpt-4o-2024-05-13| LLM, setting its ``temperature'' to zero and do not limit the number of output tokens.
We leave \coverup{}'s target code segment size (see Section~\ref{impl:code-segmentation}) at its default of 50 lines and automatically repeat test executions to look for flaky tests (see Section~\ref{impl:handling-flakies}).

We first run it without any pre-existing test suite to place \coverup{} on the same footing as \codamosa{}.
We then run it twice more, allowing it to build upon the test suite from the previous runs.
Since GPT-generated tests often assume the presence of certain Python modules, we configure \coverup{} to install any such missing modules automatically.

\subsection{[RQ1] Comparison to the previous state of the art}\label{evaluation:comparison}
To compare \coverup{} to the previous state of the art, we first evaluate it on the CM suite, using \codamosa~(codex) and \codamosa~(gpt4o) as baselines.
Figure~\ref{fig:cm-coverage} shows, on the left, the line, branch, and combined line and branch coverage obtained by \coverup{} and the baselines over the entire benchmark suite; on the right, it shows the median of those measurements on a module-by-module basis.
Additionally, Figure~\ref{fig:comparison} plots the difference between the combined line and branch coverage obtained by \coverup{} and \codamosa{} on that suite, also on a module-by-module basis; green bars in the plot indicate modules where \coverup{} achieved higher coverage.

\iffigreorg
\begin{figure}
    \centering
    \captionsetup[subfigure]{labelformat=empty}
    \begin{subfigure}[t]{0.5\textwidth}
        \includegraphics[width=\columnwidth]{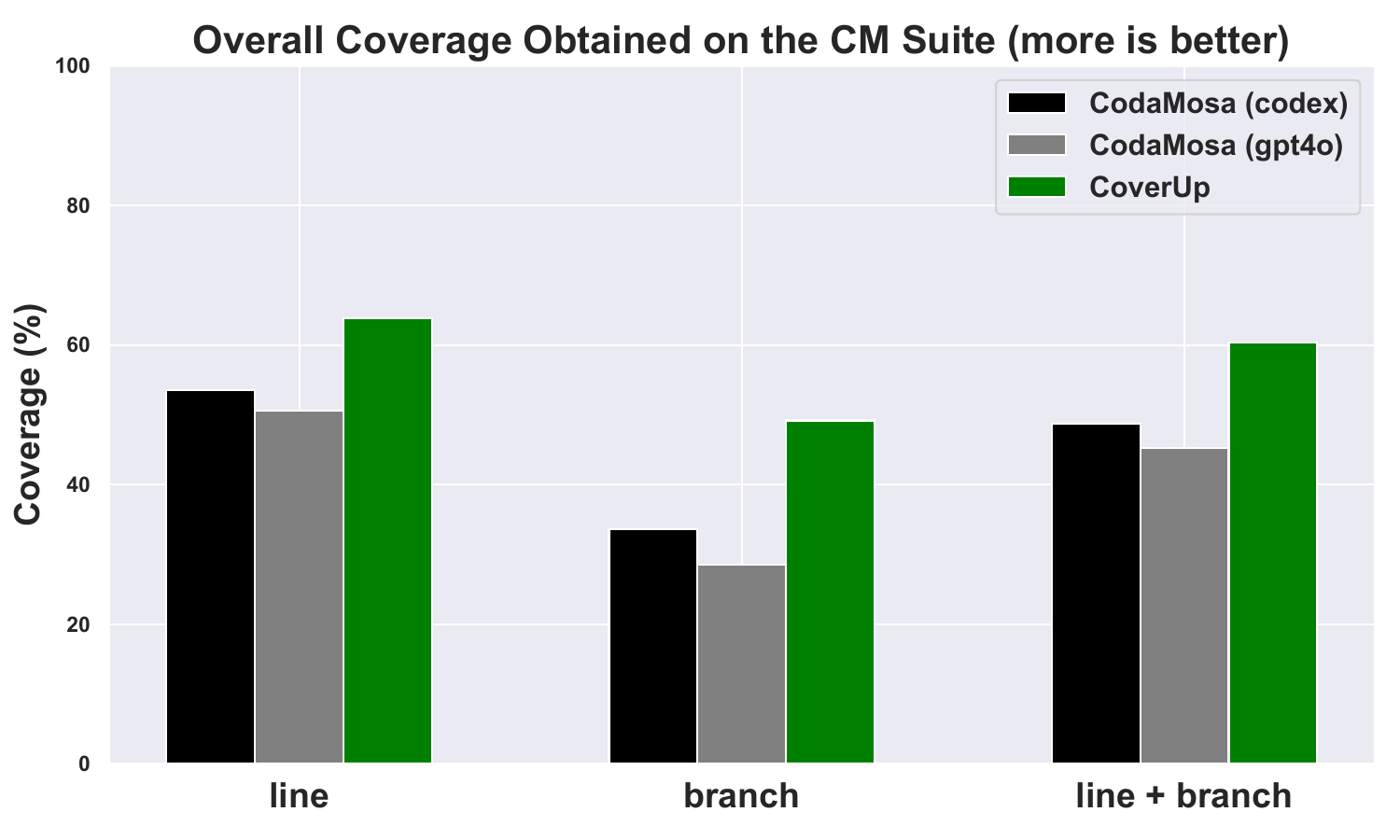}
    \end{subfigure}%
    ~\hfill~
    \begin{subfigure}[t]{0.5\textwidth}
        \includegraphics[width=\columnwidth]{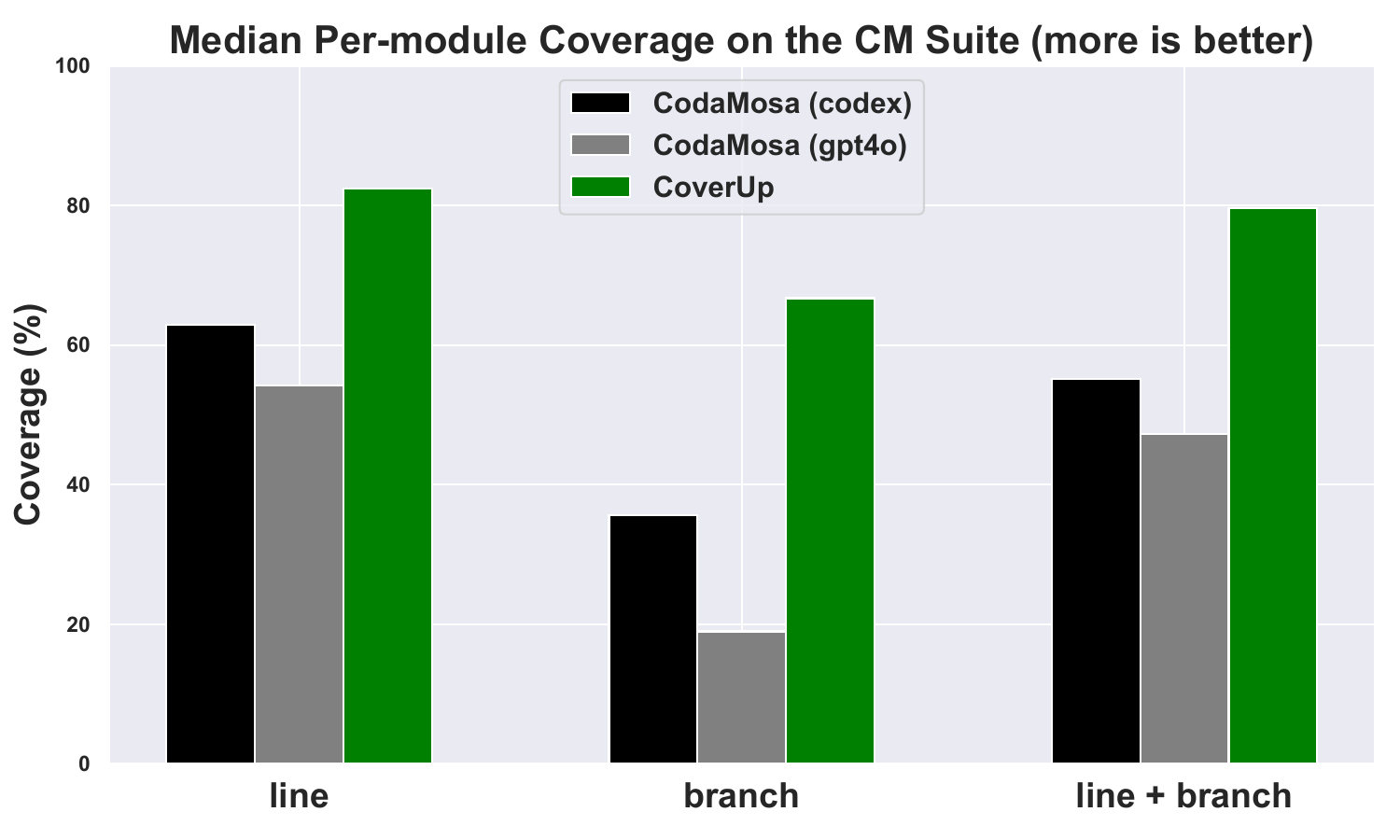}
    \end{subfigure}
    \caption{\textbf{[RQ1] \coverup{} yields higher overall and median per-module coverage:}
        \textnormal{
            Across the board, \coverup{} yields higher coverage than \codamosa{}, whether measured over the entire suite or on a module-by-module basis.
        }
        \label{fig:cm-coverage}
    }
\end{figure}

\begin{figure}[]
    \begin{minipage}[t]{0.46\textwidth}
        \input{fig-cm-comparison}
    \end{minipage}
    \hspace{15pt}
    \begin{minipage}[t]{0.46\textwidth}
        \input{fig-py-suite-content}
    \end{minipage}
\end{figure}
\begin{figure}
    \centering
    \captionsetup[subfigure]{labelformat=empty}
    \begin{subfigure}[t]{0.5\textwidth}
        \includegraphics[width=\columnwidth]{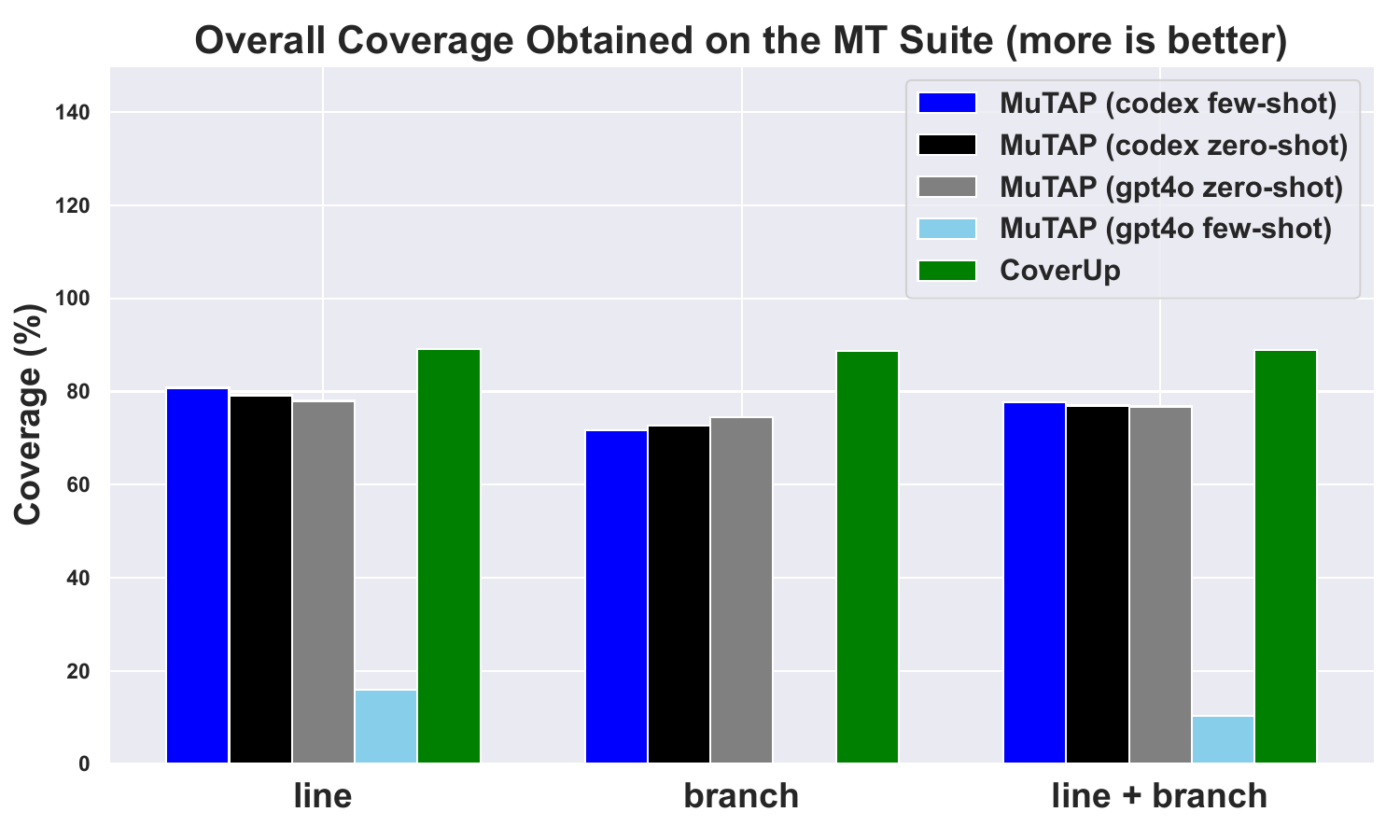}
    \end{subfigure}%
    ~\hfill~
    \begin{subfigure}[t]{0.5\textwidth}
        \includegraphics[width=\columnwidth]{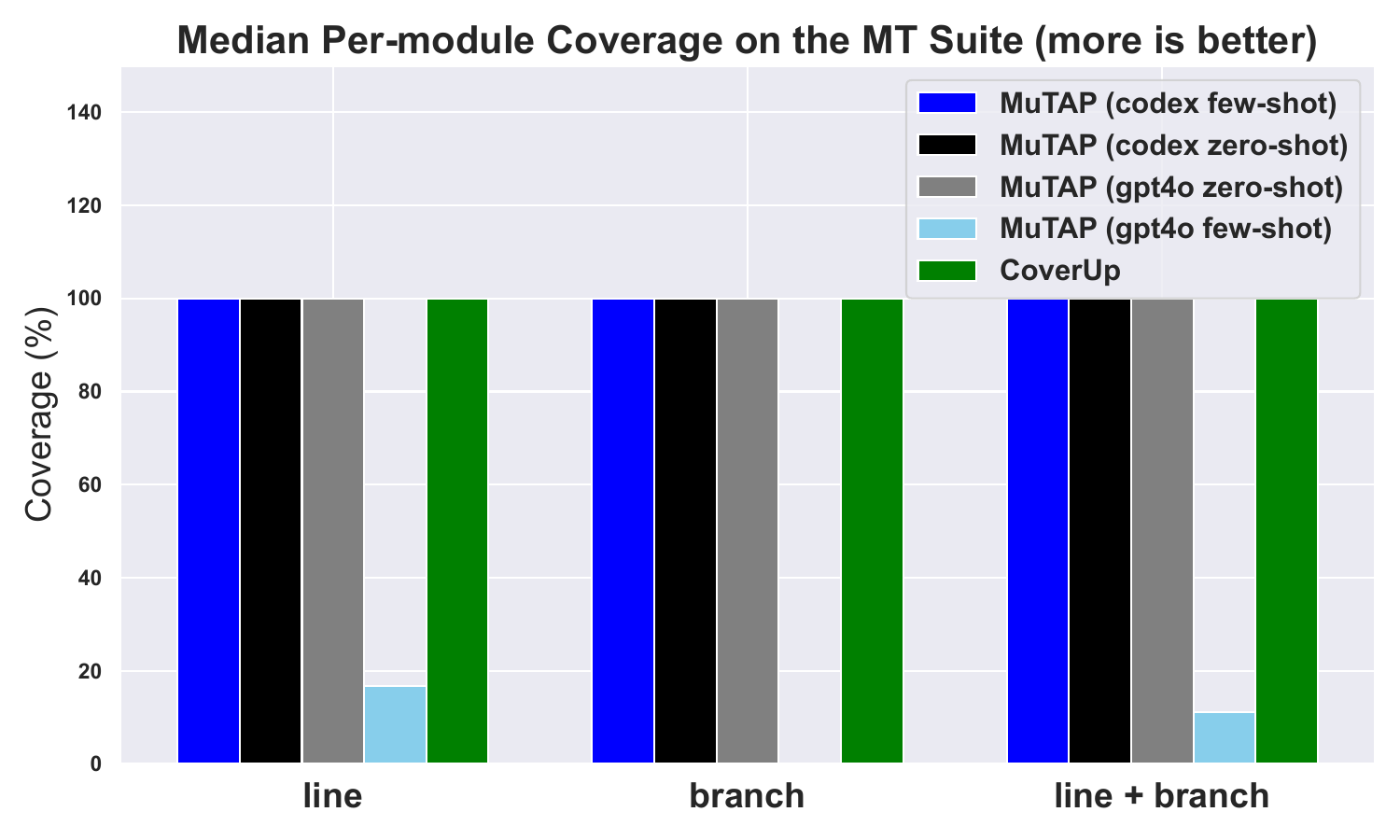}
    \end{subfigure}
    \caption{\textbf{[RQ1] \coverup{} yields higher overall and per-module coverage:}
        \textnormal{
            \coverup{} yields higher or equal coverage than MuTAP, whether measured over the entire suite or on a module-by-module basis.
        }
        \label{fig:cm-coverage-mutap}
    }
\end{figure}

\fi

As the figures show, \coverup{} achieves higher line, branch, and combined line and branch coverage than both \codamosa{} baselines, both measuring over the entire benchmark code base and on a per-module basis.
Across the entire benchmark suite, \coverup{} achieves 64\% (vs. 54\% and 51\%) line coverage, 49\% (vs. 34\% and 29\%) branch coverage, and 60\% (vs. 49\% and 45\%) line+branch coverage.
On a per-module basis, \coverup{} achieves 82\% (vs. 63\% and 54\%) line coverage, 67\% (vs. 36\% and 19\%) branch coverage, and 80\% (vs. 55\% and 47\%) line+branch coverage.
These per-module improvements over CodaMosa~(GPT-4o) and CodaMosa~(Codex) are statistically significant: using paired permutation tests, we obtain a $p$-value of $2.0 \times 10^{-5}$ for both, well below the standard $p<0.05$ threshold.
We also observe that \codamosa~(gpt4o)'s performance falls slightly behind that of \codamosa~(codex), with overall coverage measurements within 5\% of each other.
As this difference shows, a newer model does not necessarily lead to higher performance.

As Figure~\ref{fig:comparison} shows, \coverup{} does not achieve higher coverage than
\codamosa{} for some modules.
We examine \coverup{}'s logs and observe that timeouts running tests for modules in the \texttt{youtube\_dl} package constitute the largest single source of failure.
While the authors of \codamosa{} also report problems with timeouts for that package~\cite{codamosa}, it seems likely that \coverup{}'s test execution timeout of one minute was set too low.
In other cases, we observe that \coverup{}'s static analysis cannot provide the LLM with correct context when the code under test contains conditional imports, such as those intended to accommodate package differences across Python versions.
Additionally, while LLM test generations often include multiple test functions in the same response, \coverup{} rejects them all if any of them fails; it is possible that \coverup{}'s performance would have been higher if it were to evaluate each test function individually.

Next, we evaluate the performance of \coverup{} on code for which the original Pynguin performed well.
Figure~\ref{fig:py-suite} shows the results.
Given \coverup{}'s 100\% median per-module coverage and near 100\% overall coverage results, we conclude that \coverup{} also performs well on such code.

We then compare \coverup{} to MuTAP, evaluating it on the MT suite.
Figure~\ref{fig:cm-coverage-mutap} shows, on the left, the line, branch, and combined line and branch coverage obtained by \coverup{} and the baselines over the entire benchmark suite; on the right, it shows the median of those measurements on a module-by-module basis.
As the figures show, \coverup{} achieves higher line, branch, and combined line and branch coverage than the MuTAP baselines when measuring over the entire benchmark and greater or equal coverage when measuring on a per-module basis.
Across the entire benchmark suite, \coverup{} achieves 89\% (vs. 81\% to 16\%) line coverage, 89\% (vs. 73\% to 0\%) branch coverage, and 89\% (vs. 78\% to 10\%) line+branch coverage; on a per-module basis, like its baselines, \coverup{} achieves 100\% on all metrics, except for MuTAP~(gtp4o few-shot), which achieves extremely low coverage (17\%, 0\% and 11\%).

We investigate MuTAP~(gpt4o few-shot)'s extremely low performance and discover that its original few-shot prompt confuses GPT-4o: it generates tests for the test generation examples included in that prompt rather than for the function under test.
When this happens, even if the resulting tests run to completion, they do not contribute to coverage.

As the consistently high per-module median coverage values indicate, the MT suite does not contain particularly challenging code.
In fact, the functions in the suite are entirely self-contained.
They only rarely utilize external code; when they do so, they use it from standard libraries.
Additionally, the functions commonly include type annotations.
All of these characteristics greatly simplify the task of test generation.
By contrast, the functions in the CM suite have numerous dependencies, often use external code, and are generally unannotated.


\iffigreorg\else
\begin{figure}
    \centering
    \captionsetup[subfigure]{labelformat=empty}
    \begin{subfigure}[t]{0.5\textwidth}
        \includegraphics[width=\columnwidth]{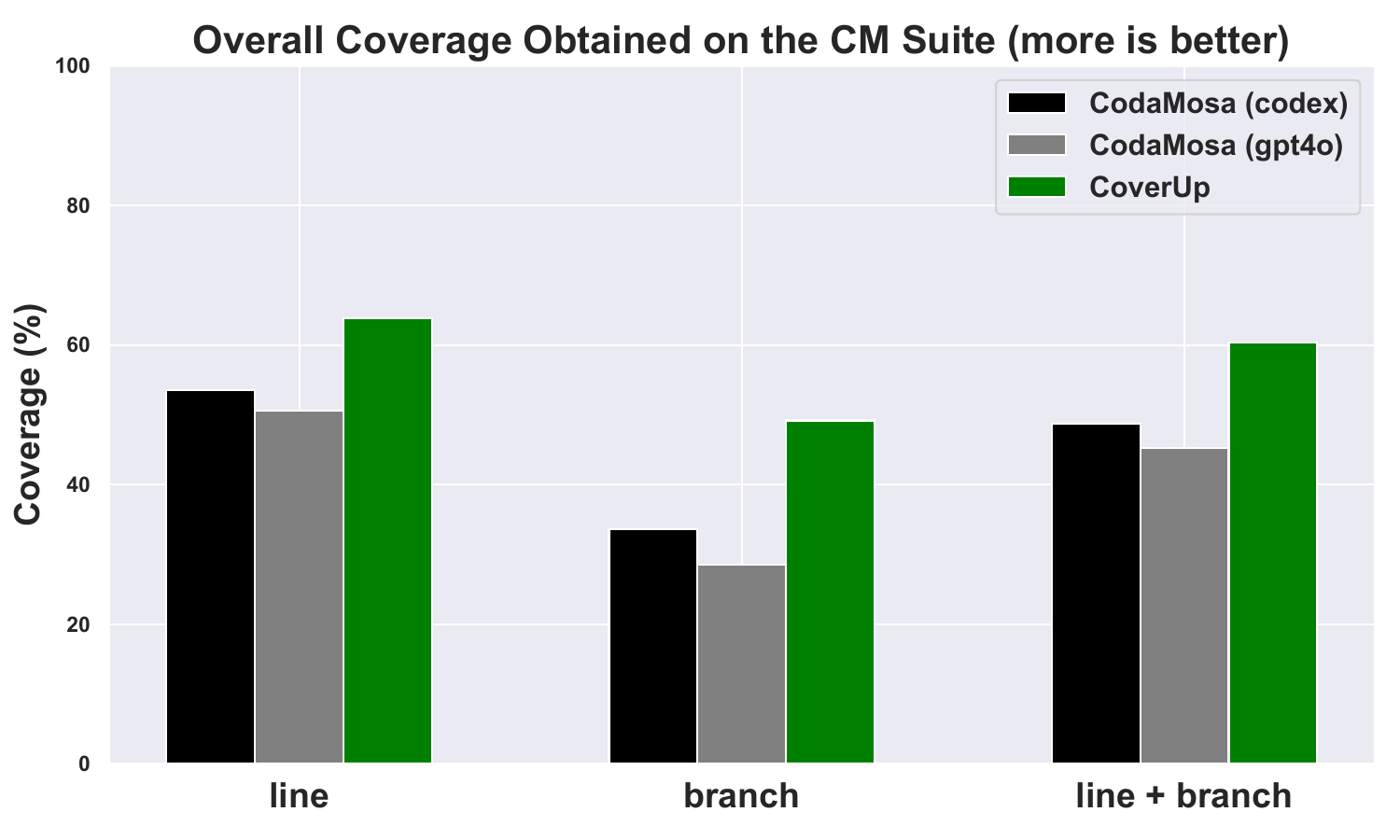}
    \end{subfigure}%
    ~\hfill~
    \begin{subfigure}[t]{0.5\textwidth}
        \includegraphics[width=\columnwidth]{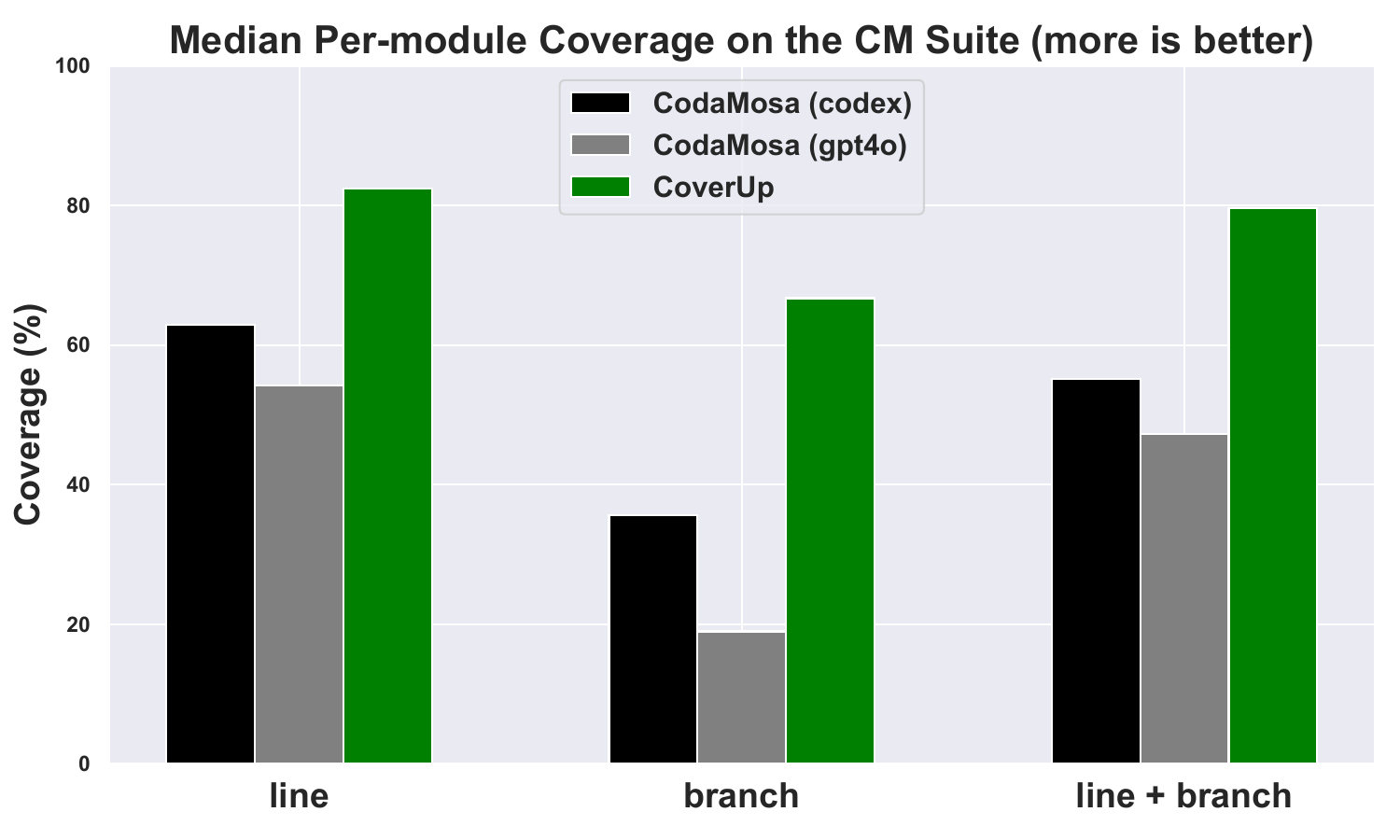}
    \end{subfigure}
    \caption{\textbf{[RQ1] \coverup{} yields higher overall and median per-module coverage:}
        \textnormal{
            Across the board, \coverup{} yields higher coverage than \codamosa{}, whether measured over the entire suite or on a module-by-module basis.
        }
        \label{fig:cm-coverage}
    }
\end{figure}

\begin{figure}
    \centering
    \captionsetup[subfigure]{labelformat=empty}
    \begin{subfigure}[t]{0.5\textwidth}
        \includegraphics[width=\columnwidth]{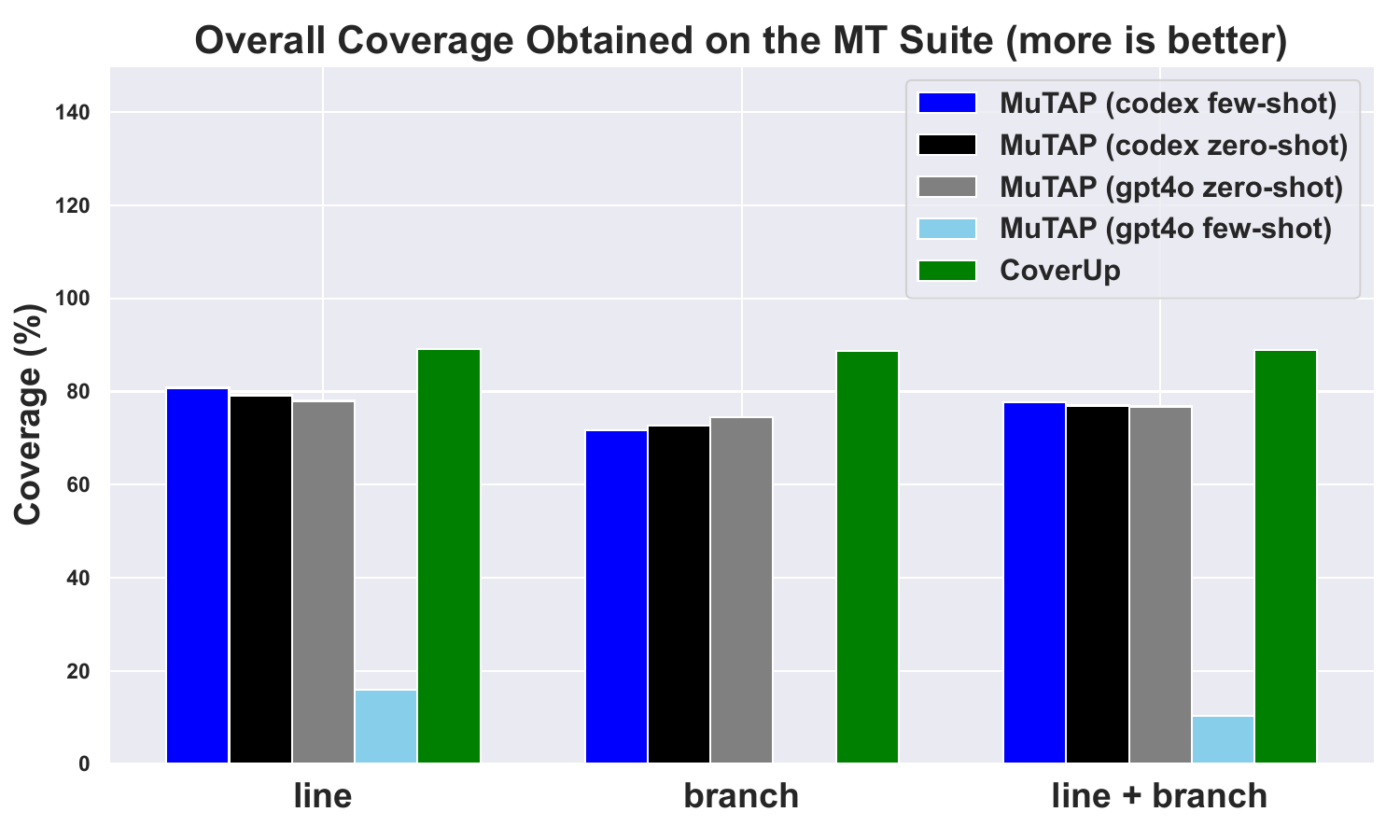}
    \end{subfigure}%
    ~\hfill~
    \begin{subfigure}[t]{0.5\textwidth}
        \includegraphics[width=\columnwidth]{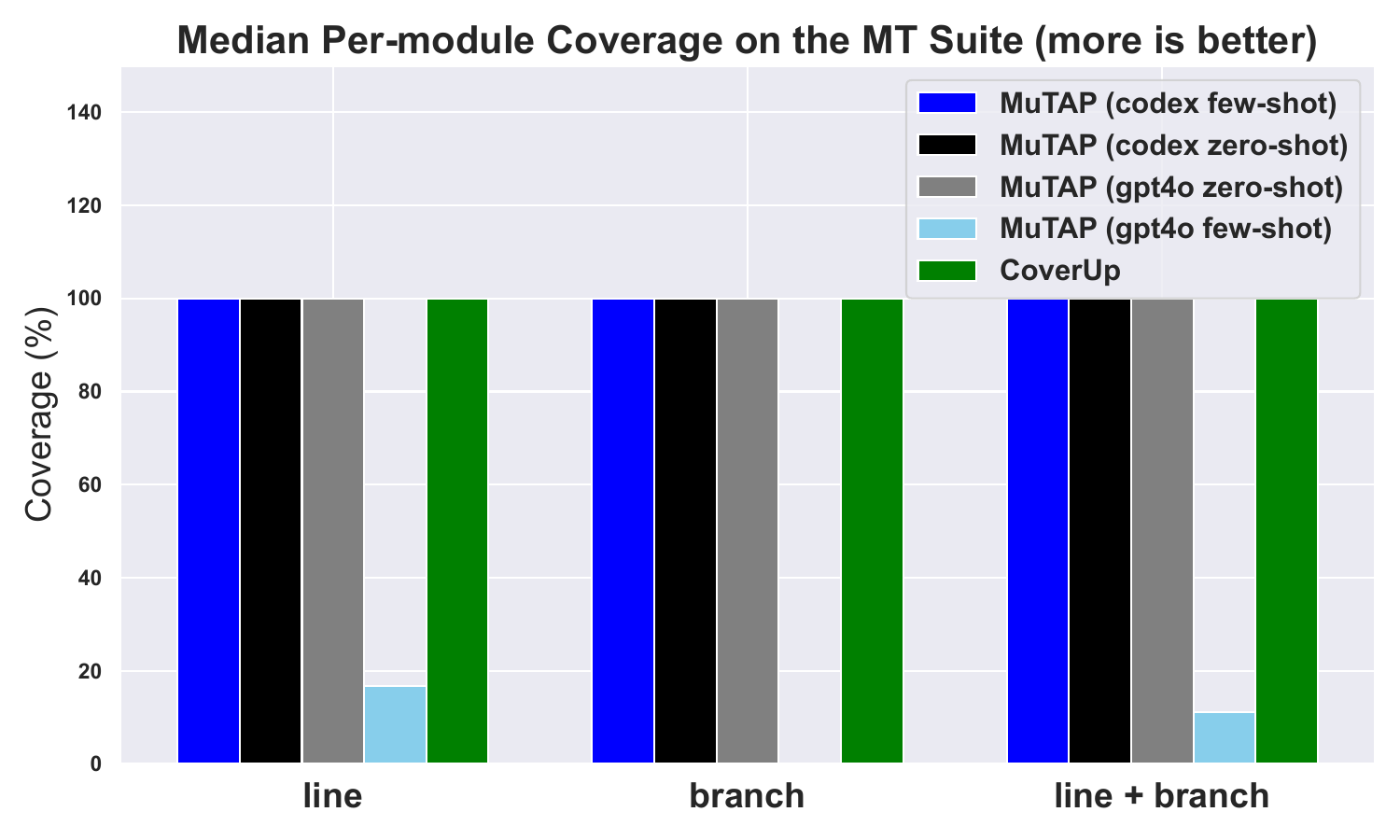}
    \end{subfigure}
    \caption{\textbf{[RQ1] \coverup{} yields higher overall and per-module coverage:}
        \textnormal{
            \coverup{} yields higher or equal coverage than MuTAP, whether measured over the entire suite or on a module-by-module basis.
        }
        \label{fig:cm-coverage-mutap}
    }
\end{figure}

\begin{figure}
    \input{fig-cm-comparison}
\end{figure}

\begin{figure}[h]
    \centering
    \begin{minipage}[t]{0.48\textwidth}
        \centering
        \input{fig-py-suite-content}
    \end{minipage}
    ~\hfill~
    \begin{minipage}[t]{0.48\textwidth}
        \input{fig-continued-prompts-content}
    \end{minipage}
\end{figure}
\fi

\newtcolorbox{conclusion}{%
    colback=green!5!white,
    colframe=green!75!black,
    arc=1mm,
    top=1mm,
    bottom=1mm,
    left=1mm,
    right=1mm
}

\begin{conclusion}
\textbf{[RQ1] Summary:} \coverup{} achieves significantly higher coverage than both \codamosa{} and MuTAP, outperforming the state of the art.
\end{conclusion}

\iffigreorg
    \begin{table}[]
    \small
    \caption{\textbf{[RQ1, RQ2] \coverup{} outperforms \codamosa{} on the CM suite} and far outperforms itself when ablated to simply rely on LLM performance (top).
    Bottom line: as the results for the PY suite show, \coverup{}'s performance is not limited to ``challenging'' code.
        \label{table:cm-coverage}
    }
    \begin{tabularx}{\columnwidth}{@{} X | r r r | r r r @{}}
        \hline
        \multicolumn{1}{l|}{\thead{\vspace{-5pt}}} & \multicolumn{3}{c|}{\thead{\vspace{-5pt}Overall Coverage}} & \multicolumn{3}{c}{\thead{\vspace{-5pt}Median Per-Module Coverage}} \\
        \thead[l]{Test Generator} & \thead[r]{Line} & \thead[r]{Branch} & \thead[r]{Line + Branch} & \thead[r]{Line} & \thead[r]{Branch} & \thead[r]{Line + Branch} \\
        \hline
        \coverup{}         & \PCT{63.8} & \PCT{49.2} & \PCT{60.3} & \PCT{82.4} & \PCT{66.7} & \PCT{79.6} \\
        \codamosa~(codex)  & \pct{53.5} & \pct{33.6} & \pct{48.7} & \pct{62.9} & \pct{35.6} & \pct{55.1} \\
        \codamosa~(gpt4o)  & \pct{50.6} & \pct{28.6} & \pct{45.2} & \pct{54.2} & \pct{19.0} & \pct{47.2} \\
        \coverup~(ablated) & \pct{38.8} & \pct{19.0} & \pct{34.0} & \pct{43.8} & \pct{7.3}  & \pct{38.5} \\
        \hline
        \coverup{} \tiny on PY suite & \pct{97.0} & \pct{99.2} & \pct{97.1} & \pct{100.0} & \pct{100.0} & \pct{100.0} \\
        \hline
    \end{tabularx}
\end{table}

    \begin{figure}
    \centering
    \captionsetup[subfigure]{labelformat=empty}
    \begin{subfigure}[t]{0.5\textwidth}
    \includegraphics[width=\columnwidth]{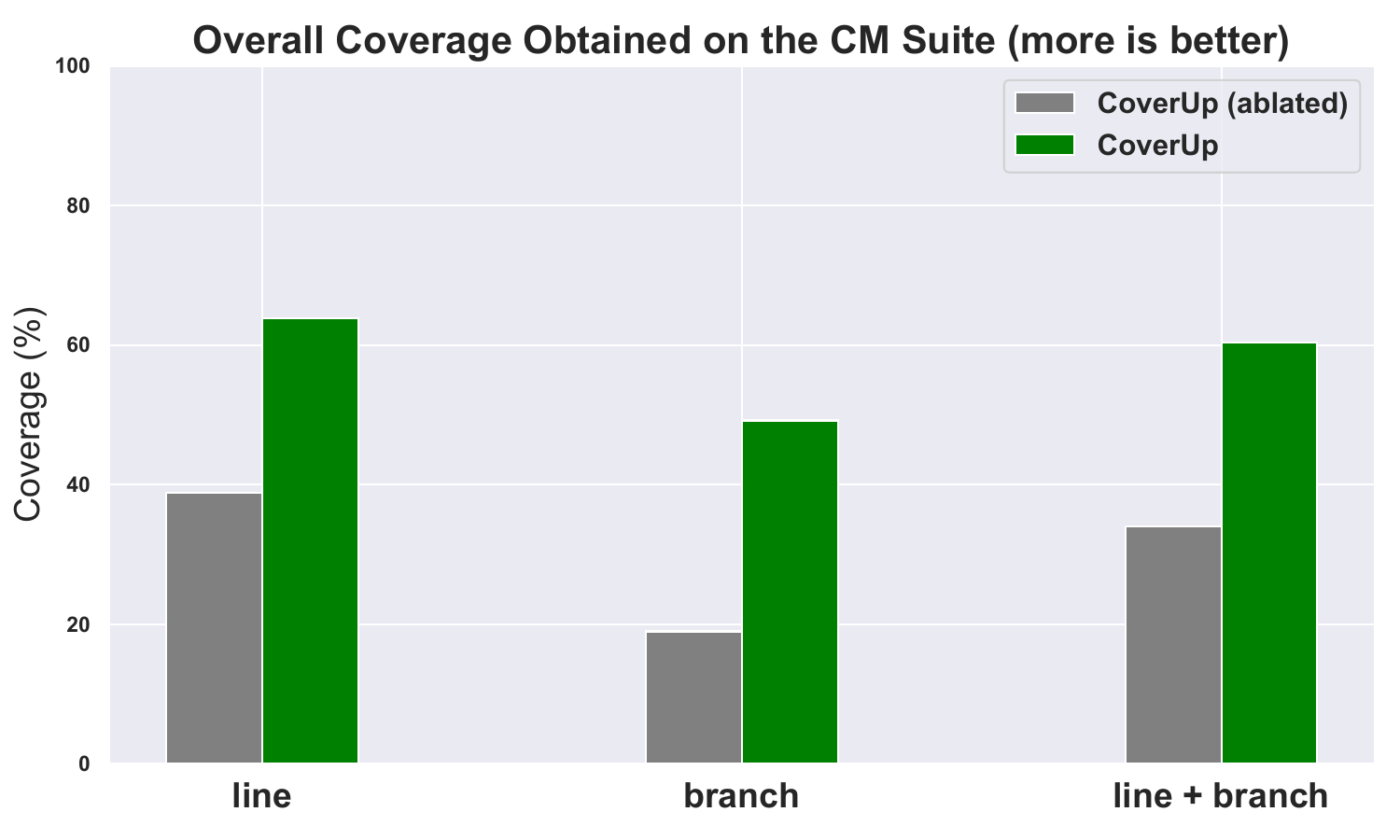}
    \end{subfigure}%
    ~
    \hfill
    ~
    \begin{subfigure}[t]{0.5\textwidth}
    \includegraphics[width=\columnwidth]{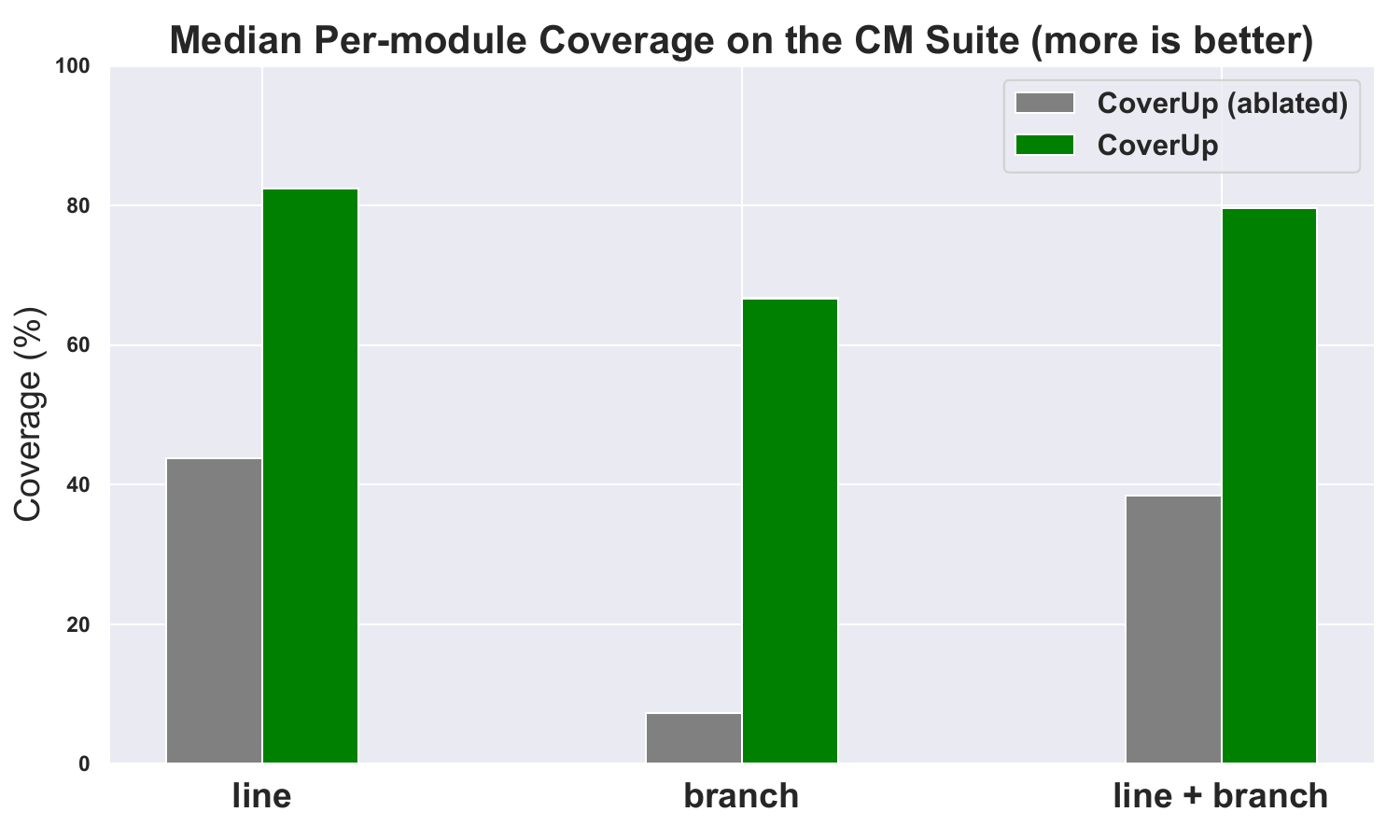}
    \end{subfigure}
    \caption{\textbf{[RQ2] \coverup{} contributes significantly to overall performance:}
        \textnormal{
            Across the board, \coverup{} yields higher coverage in comparison to the ablated (LLM only) version, whether measured over the entire suite or on a module-by-module basis.
        }
        \label{fig:cm-ablated-coverage}
    }
\end{figure}

\fi

\subsection{[RQ2] How effective is \coverup{} compared to simply prompting an LLM for tests?}
\label{evaluation:just-prompt}
Given many LLMs' near-human performance on various tasks, including software engineering tasks~\cite{bubeck2023sparks}, it is reasonable to ask just how much \coverup{} contributes to performance.
To address this question, we compare \coverup{} to an ablated LLM-only version that utilizes a nearly identical prompt but which lacks all other \coverup{} functionality (coverage information, computed \texttt{import} statements, \texttt{get\_info} tool function and continued chats).
Rather than ask for tests for the uncovered code, it asks instead for tests ``that execute \emph{all} lines and branches''.

As Figure~\ref{fig:cm-ablated-coverage} shows, \coverup{} achieves substantially higher line, branch, and combined line and branch coverage than \coverup{}~(ablated), both measured over the entire benchmark code base and on a per-module basis.
Across the entire benchmark suite, it achieves 64\% (vs. 39\%) line coverage, 49\% (vs. 19\%) branch coverage, and 60\% (vs. 34\%) line+branch coverage; on a per-module basis, it achieves 82\% (vs. 44\%) line coverage, 67\% (vs. 7\%) branch coverage, and 80\% (vs. 39\%) line+branch coverage.

\iffigreorg\else
\begin{figure}
    \centering
    \captionsetup[subfigure]{labelformat=empty}
    \begin{subfigure}[t]{0.5\textwidth}
    \includegraphics[width=\columnwidth]{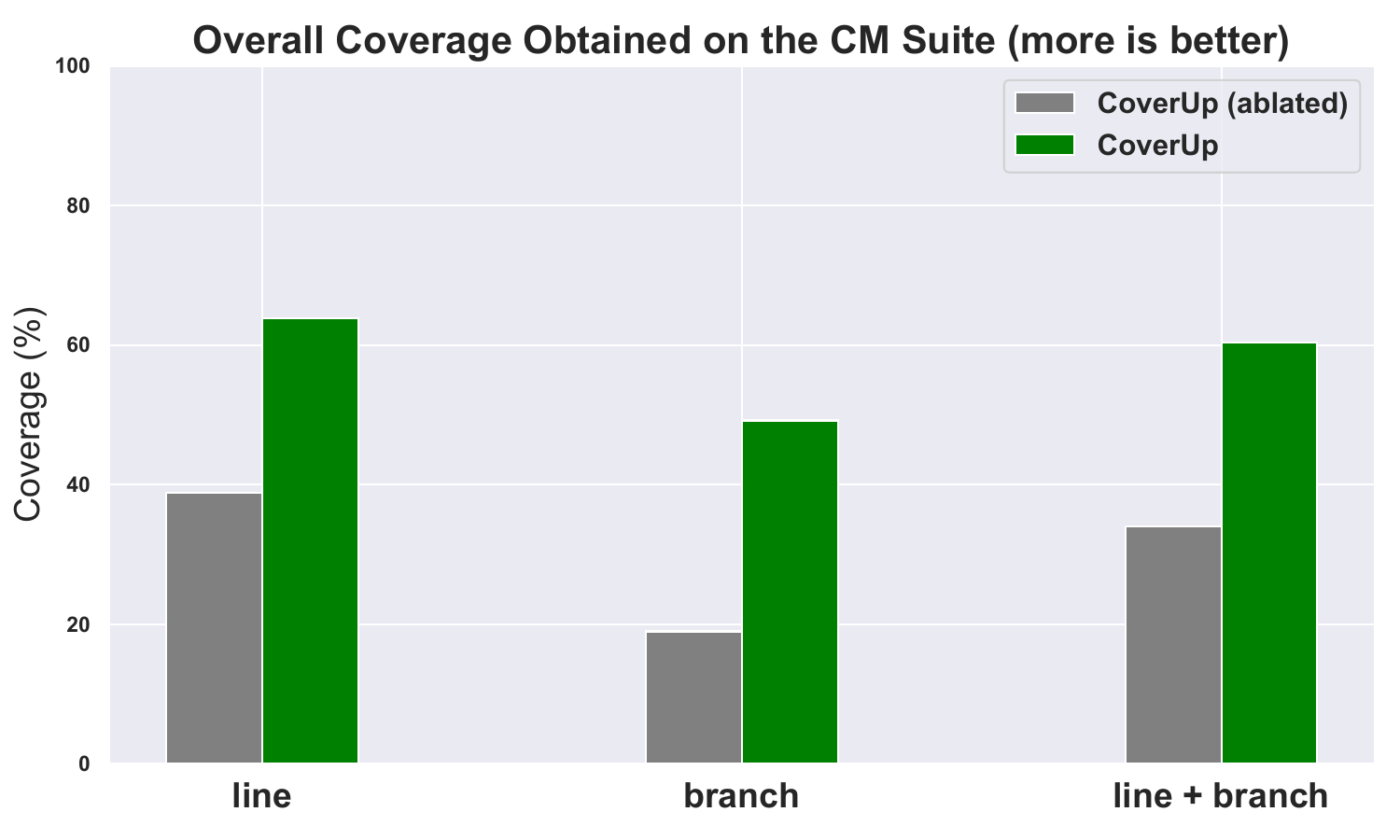}
    \end{subfigure}%
    ~
    \hfill
    ~
    \begin{subfigure}[t]{0.5\textwidth}
    \includegraphics[width=\columnwidth]{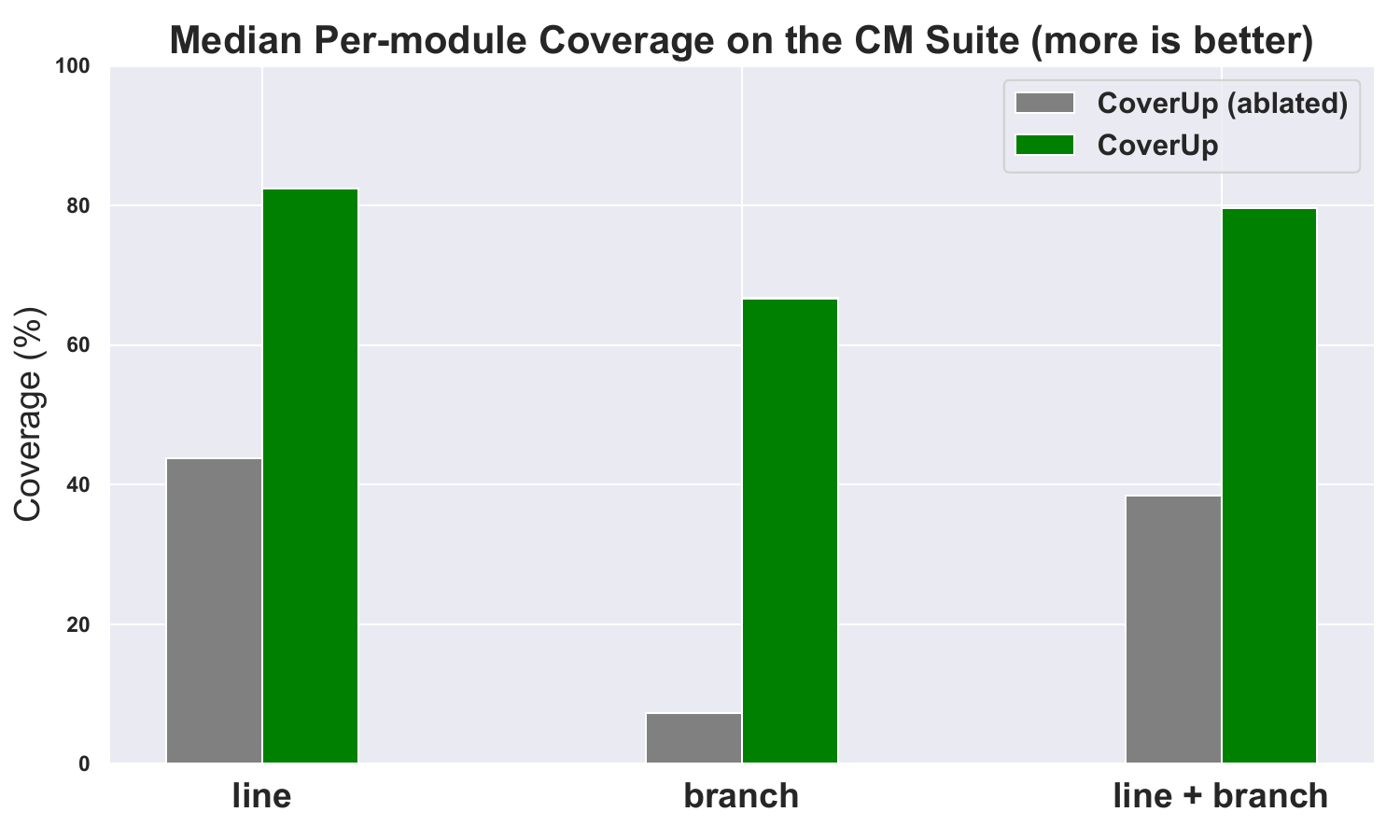}
    \end{subfigure}
    \caption{\textbf{[RQ2] \coverup{} contributes significantly to overall performance:}
        \textnormal{
            Across the board, \coverup{} yields higher coverage in comparison to the ablated (LLM only) version, whether measured over the entire suite or on a module-by-module basis.
        }
        \label{fig:cm-ablated-coverage}
    }
\end{figure}

\fi

\begin{conclusion}
\textbf{[RQ2] Summary:} \coverup{} contributes significantly to overall performance; its performance is not just due to the LLM employed.
\end{conclusion}

\iffigreorg
    \begin{table}[]
    \small
    \caption{\textbf{[RQ1] \coverup{} outperforms MuTAP's coverage on the MT suite}.
        \label{table:mutap-overall-coverage}
    }
    \begin{tabularx}{\columnwidth}{@{} X | r r r | r r r @{}}
        \hline
        \multicolumn{1}{l|}{\thead{\vspace{-5pt}}} & \multicolumn{3}{c|}{\thead{\vspace{-5pt}Overall Coverage}} & \multicolumn{3}{c}{\thead{\vspace{-5pt}Median Per-Module Coverage}} \\
        \thead[l]{Test Generator} & \thead[r]{Line} & \thead[r]{Branch} & \thead[r]{Line + Branch} & \thead[r]{Line} & \thead[r]{Branch} & \thead[r]{Line + Branch} \\
        \hline
        \coverup{}              & \PCT{89.2} & \PCT{88.7} & \PCT{89.0} & \PCT{100.0} & \PCT{100.0} & \PCT{100.0} \\
        MuTAP~(codex few-shot)  & \pct{80.8} & \pct{71.7} & \pct{77.7} & \pct{100.0} & \pct{100.0} & \pct{100.0} \\
        MuTAP~(codex zero-shot) & \pct{79.2} & \pct{72.7} & \pct{77.0} & \pct{100.0} & \pct{100.0} & \pct{100.0} \\
        MuTAP~(gpt4o zero-shot) & \pct{78.0} & \pct{74.5} & \pct{76.8} & \pct{100.0} & \pct{100.0} & \pct{100.0} \\
        MuTAP~(gpt4o few-shot)  & \pct{15.9} & \pct{0.0}  & \pct{10.4} & \pct{16.7}  & \pct{0.0}   & \pct{11.1} \\
        \hline
    \end{tabularx}
\end{table}

\fi

\subsection{[RQ3] How effective are \coverup{}'s continued dialogues?}\label{evaluation:continued-dialogue}
To investigate RQ3, we process \coverup{}'s logs generated while evaluated on the CM suite, identifying each successful test (i.e., which passes and improves coverage) that was generated immediately upon the first prompt or after continuing the chat with a second or third prompt (by default, and also in our evaluation, \coverup{} continues the conversation for at most two additional prompts).
We observe that 60.3\% of successes result from the first prompt, 27.2\% from the second, and 12.5\% from the third.
While the success rate decreases with each iteration, approximately 40\% of successes were achieved through iterative refinement of the prompt, highlighting its importance.
\begin{conclusion}
\textbf{[RQ3] Summary:} Continuing the chat contributes about 40\% of its successes, demonstrating its effectiveness.
\end{conclusion}

\subsection{[RQ4] How does the cost of running \coverup{} compare to \codamosa{}?}\label{evaluation:cost}
Examining the logs from runs on the CM suite, we assemble Table~\ref{table:cost}, which shows the number of prompts, completions, tokens, running time, and approximate cost incurred.

\begin{table}
    \small
    \caption{\textbf{[RQ4] Cost of running \coverup{} and \codamosa{} on the CM suite:}
        \coverup{} runs roughly 18$\times$ faster while achieving higher coverage
        than \codamosa{}, but uses 48\% more tokens.
        \label{table:cost}
    }
    \begin{tabularx}{\columnwidth}{@{}X | r r | r r | r | r@{}}
        \hline
        \thead[l]{Test Generator} & \thead[r]{Prompts} & \thead[r]{P. Tokens} & \thead[r]{Completions} & \thead[r]{C. Tokens} & \thead[r]{Time (h)} & \thead[r]{Cost (US\$)} \\
        \hline
        \coverup           & 28,690 & 58,081,908 & 28,654 & 7,257,400 & 4.0  & 399 \\
        \coverup~(ablated) &  9,224 &  3,501,827 &  9,188 & 3,859,444 & 1.7  & 75 \\
        \codamosa~(gpt4o)  & 30,368 & 39,508,559 & 29,136 & 4,504,329 & 71.0 & 265 \\
        \hline
    \end{tabularx}
\end{table}

We observe that \coverup{} runs roughly 18 times faster (4 vs. 71 hours) while achieving higher coverage than \codamosa{}, but using 48\% more tokens.
Although this time comparison can provide insight into each method's applicability for a given context, it is important to understand its limitations.
\coverup{} and \codamosa{} utilize both local and cloud-based computing resources; we describe the local system in Section~\ref{eval:execution-environment}, but OpenAI does not disclose details of their computing environment, whose availability may vary over time.
\coverup{}'s implementation is both parallelized and asynchronous.
Its running time is primarily gated by the rate limits imposed by OpenAI (see Section~\ref{impl:technical-challenges}), their response times, and the time required to execute the generated tests.
In contrast, \codamosa{} is a sequential process that runs for 10 minutes per module.
While this duration is configurable, we follow the setting used in the \codamosa{} paper~\cite{codamosa}.

\begin{conclusion}
\textbf{[RQ4] Summary:} As deployed, \coverup{} runs 18 times faster while achieving higher coverage than \codamosa{}, but using almost 50\% more tokens.
\end{conclusion}

\subsection{[RQ5] How important are \coverup{}'s components to its performance?}\label{evaluation:component-ablations}
In broad terms, \coverup{} contains components that perform three main functions: provide the LLM information on coverage, provide it with additional code context, and provide it with feedback on any errors.
In this research question, we explore how ablating each one of these components affects \coverup{}.
Table~\ref{table:ablated-components} shows the specific elements disabled for each variant; disabling all three yields the same ``\coverup{}~(ablated)'' system explored already in RQ2.

\newcommand{\CM}{%
\tikz[scale=0.23] {
    \draw[color=teal, line width=0.7,line cap=round] (0.25,0) to [bend left=10] (1,1);
    \draw[color=teal, line width=0.8,line cap=round] (0,0.35) to [bend right=1] (0.23,0);
}}
\newcommand{\XM}{%
\tikz[scale=0.23] {
    \draw[color=red, line width=0.7,line cap=round] (0,0) to [bend left=6] (1,1);
    \draw[color=red, line width=0.7,line cap=round] (0.2,0.95) to [bend right=3] (0.8,0.05);
}}
\settowidth\rotheadsize{\theadfont no code context}
\renewcommand\theadalign{bc}
\begin{table}[]
    \small
    \caption{\textbf{[RQ5] Elements in \coverup{} ablations:} The non-ablated \coverup{} contains all of these elements, while \coverup~(ablated) from RQ2 contains none of them. To answer RQ5, we create three additional ablations.
        \label{table:ablated-components}
    }
    \begin{tabularx}{\columnwidth}{@{} X | c c c c c @{}}
        \thead[l]{Component} & \rothead{non-ablated} & \rothead{ablated [RQ2]} & \rothead{no coverage} & \rothead{no code context} & \rothead{no error fixing} \\
        \hline
        Indication of what coverage is missing                      &  \CM  & \XM  & \XM  &  \CM  &  \CM  \\
        Continued chat if generated tests do not improve coverage   &  \CM  & \XM  & \XM  &  \CM  &  \CM  \\
        Generated \texttt{import} statements                        &  \CM  & \XM  & \CM  &  \XM  &  \CM  \\
        The \texttt{get\_info} tool function                        &  \CM  & \XM  & \CM  &  \XM  &  \CM  \\
        Continued chat in case of errors                            &  \CM  & \XM  & \CM  &  \CM  &  \XM  \\
        \hline
    \end{tabularx}
\end{table}
\begin{figure}[t]
    \centering
    \begin{minipage}[t]{0.48\textwidth}
        \centering
        \includegraphics[width=\linewidth]{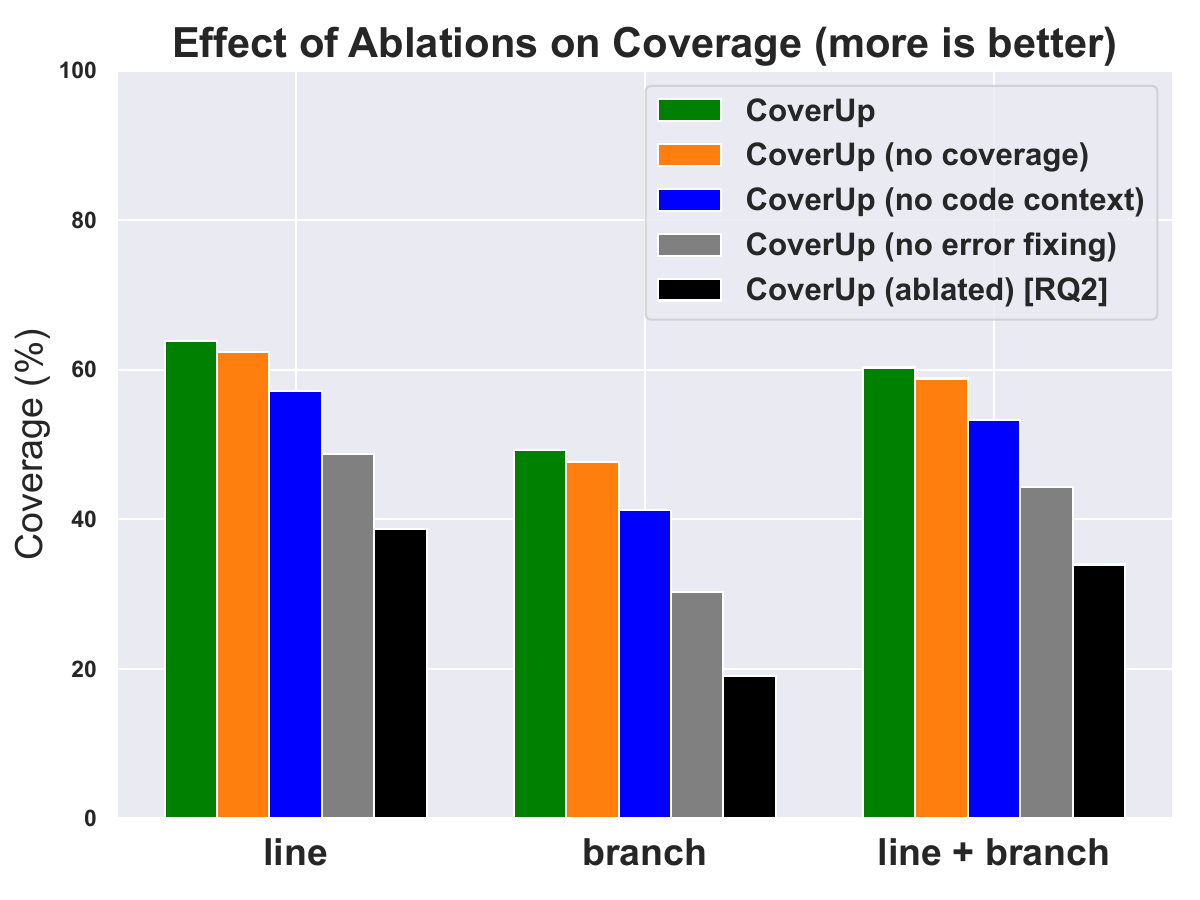}
        \caption{\textbf{[RQ5] Effect of ablations on coverage:} Removing any of these components from \coverup{} lowers its performance.
            \label{fig:component-ablation-results}
        }

    \end{minipage}
    \hspace{10pt}
    \begin{minipage}[t]{0.48\textwidth}
        \centering
        \includegraphics[width=\linewidth]{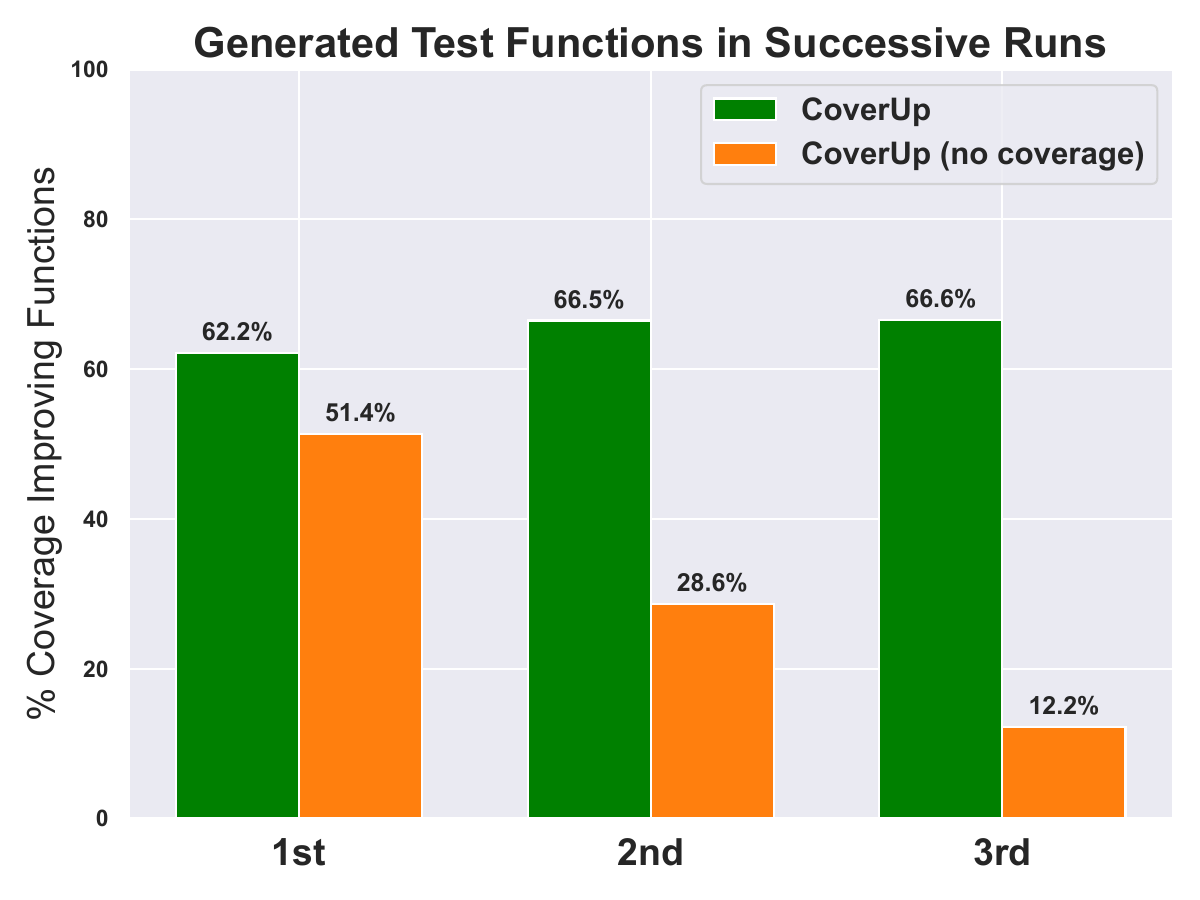}
        \caption{\textbf{[RQ5] Importance of coverage-based prompting:} As the starting coverage grows with each successive run, the coverage-ablated prompt generates fewer and fewer functions that increase coverage.
            \label{fig:importance-of-coverage-prompting}
        }
    \end{minipage}
\end{figure}
We first evaluate these new ablations of \coverup{} using suite CM; Figure~\ref{fig:component-ablation-results} shows the resulting overall coverage for each ablation.
We observe that all ablations negatively affect coverage, indicating their importance.
Ablating error fixing has the most significant effect, losing between 14\% and 37\% on the various coverage metrics; ablating code context loses between 7\% and 16\%, and ablating coverage information up to 2\%.
In interpreting these results, it is important to realize that the components are not entirely independent.
For example, the feedback from error fixing may allow the LLM to correct for insufficient code context, and through improved code context, errors resulting from incorrect assumptions may be avoided.

To better understand the effect of ablating coverage information from the prompt, we count the test functions generated by the model and assess how many of these increase coverage for each successive run.
As Figure~\ref{fig:importance-of-coverage-prompting} shows, as coverage increases, the coverage-ablated prompt results in fewer and fewer functions that increase coverage.
In contrast, \coverup{}'s prompt remains effective.
This result shows that, by providing coverage information in the prompt, \coverup{} focuses the LLM's attention on uncovered code.
Even though the coverage-ablated prompt only fell short in coverage by a small amount, it did so by generating over 50\% more functions than the non-ablated prompt (11,222 vs. 7,366).
If left in the test suite, these functions contribute to bloat, slowing the suite's execution.
But even if rejected after a coverage check, they add unnecessary tokens in LLM responses and thus increase cost.
\ifarxiv

This analysis also suggests potential improvements for \coverup{}.  
Since even the non-ablated prompt results in some test functions that do not increase coverage, \coverup{} should measure the coverage generated by each function individually, rather than for the entire LLM response.  
Additionally, \coverup{} could omit coverage information when prompting for tests for entirely uncovered functions, thereby reducing the cost of tagging every line in a code segment.
We leave this to future work.
\fi
\ifarxiv
\begin{table}[]
    \small
    \caption{\textbf{[RQ5] Effect of ablations on coverage:} Removing any of these components from \coverup{} lowers its performance.
        \label{table:component-ablation-results}
    }
    \begin{tabularx}{\columnwidth}{@{} X | r r r | r r r @{}}
        \hline
        \multicolumn{1}{l|}{\thead{\vspace{-5pt}}} & \multicolumn{3}{c|}{\thead{\vspace{-5pt}Overall Coverage}} & \multicolumn{3}{c}{\thead{\vspace{-5pt}Median Per-Module Coverage}} \\
        \thead[l]{Test Generator} & \thead[r]{Line} & \thead[r]{Branch} & \thead[r]{Line + Branch} & \thead[r]{Line} & \thead[r]{Branch} & \thead[r]{Line + Branch} \\
        \hline
        \coverup~(no~coverage)           & \pct{-1.4} & \pct{-1.6} & \pct{-1.5} & \pct{-1.2} & \pct{-1.9} & \pct{-0.3} \\
        \coverup~(no~code~context)       & \pct{-6.7} & \pct{-8.0} & \pct{-7.0} & \pct{-8.4} & \pct{-15.6} & \pct{-9.1} \\
        \coverup~(no~error~fixing)       & \pct{-15.1} & \pct{-18.9} & \pct{-16.0} & \pct{-24.6} & \pct{-37.1} & \pct{-27.9} \\
        \coverup~(ablated) [RQ2]         & \pct{-25.1} & \pct{-30.2} & \pct{-26.3} & \pct{-38.6} & \pct{-59.4} & \pct{-41.0} \\
        \hline
    \end{tabularx}
\end{table}
\begin{table}[]
    \small
    \caption{\textbf{[RQ5] Importance of coverage-based prompting:} As the starting coverage grows with each successive run, the coverage-ablated prompt generates fewer and fewer functions that increase coverage.
        \label{table:importance-of-coverage-prompting}
    }
    \begin{tabularx}{\columnwidth}{@{} X | r r | r r | r r | r r @{}}
        \hline
        \multicolumn{1}{l|}{\thead{\vspace{-5pt}}} & \multicolumn{2}{c|}{\thead{\vspace{-5pt}First Run}} & \multicolumn{2}{c|}{\thead{\vspace{-5pt}Second Run}} & \multicolumn{2}{c|}{\thead{\vspace{-5pt}Third Run}} & \multicolumn{2}{c}{\thead{\vspace{-5pt}End}} \\
        \thead[l]{Test Generator}
        & \thead[r]{Initial\\Cov.} & \thead[r]{Cov. Incr.\\Functions}
        & \thead[r]{Initial\\Cov.} & \thead[r]{Cov. Incr.\\Functions}
        & \thead[r]{Initial\\Cov.} & \thead[r]{Cov. Incr.\\Functions}
        & \thead[r]{Cov.} & \thead[r]{Total\\Functions}
        \\
        \hline
        \coverup                 & \pct{0.0} & \pct{62.2} & \pct{49.5} & \pct{66.5} & \pct{56.7} & \pct{66.6} & \pct{60.3} & 7,366 \\
        \coverup~(no~coverage)   & \pct{0.0} & \PCT{51.4} & \pct{51.0} & \PCT{28.6} & \pct{56.9} & \PCT{12.2} & \pct{58.8} & 11,222 \\
        \hline
    \end{tabularx}
\end{table}
\fi
\begin{conclusion}
\textbf{[RQ5] Summary:} Coverage-based prompting, code context, and error fixing are all important components of \coverup{}; ablating any of them results in reduced coverage.
\end{conclusion}

    \section{Threats to Validity}

\subsubsection*{Benchmark selection}
We utilize the CM and PY benchmark suites to evaluate \coverup{}'s performance on both challenging and less challenging code, as well as to facilitate including Codex-based results; we utilize the MT benchmark suite because MuTAP supports it, and it, too, facilitates including Codex-based results.
While our experience executing \coverup{} to generate tests for other software has yielded similarly high coverage, selecting a different set of benchmarks could produce different results.

\subsubsection*{Execution environment}
\codamosa{}'s original evaluation environment failed to install a number of Python modules that are prerequisites for the applications used as benchmarks.
In an effort to replicate the original conditions as closely as possible, we ignored these failures as well.
It seems likely that both \codamosa{} and \coverup{} would be better able to generate tests if these modules were not missing.

\subsubsection*{LLM model dependency}
\coverup{} was developed and evaluated using OpenAI's GPT-4, GPT-4 Turbo and GPT-4o models.
\coverup{}'s approach is independent of the LLM and, as Section~\ref{evaluation:just-prompt} shows, \coverup{} significantly outperforms an ablated version of itself that just relies on the LLM's capabilities.
Its ultimate performance naturally depends on the model's ability to interpret its prompts and generate tests as requested.

    \section{Conclusion}
This paper introduces \coverup{}, a novel approach to guide LLM-based test generation through coverage analysis.  
While modern LLMs achieve near-human performance on various tasks, simply prompting them is insufficient for generating high-coverage tests.  
We demonstrate that integrating coverage information into prompts directs the LLM’s attention to uncovered code and highlight the benefits of iterative refinement through error and coverage feedback.  
By combining these elements with code context, \coverup{} produces test suites that achieve higher coverage than previous approaches.  

\section{Data Availability}
\ifanonymous
\coverup{} sources are available anonymized at \coverupurl{} and \cleanslateurl{}.
We intend to make data publicly available upon acceptance.
\else
A replication package is available at \replicationurl{};\linebreak
\coverup{} is available on GitHub at \coverupurl{} and archived on Zenodo~\cite{zenodo_coverup, zenodo_cleanslate}.
\fi

    \iffigreorg\else
        \begin{table}[]
    \small
    \caption{\textbf{[RQ1, RQ2] \coverup{} outperforms \codamosa{} on the CM suite} and far outperforms itself when ablated to simply rely on LLM performance (top).
    Bottom line: as the results for the PY suite show, \coverup{}'s performance is not limited to ``challenging'' code.
        \label{table:cm-coverage}
    }
    \begin{tabularx}{\columnwidth}{@{} X | r r r | r r r @{}}
        \hline
        \multicolumn{1}{l|}{\thead{\vspace{-5pt}}} & \multicolumn{3}{c|}{\thead{\vspace{-5pt}Overall Coverage}} & \multicolumn{3}{c}{\thead{\vspace{-5pt}Median Per-Module Coverage}} \\
        \thead[l]{Test Generator} & \thead[r]{Line} & \thead[r]{Branch} & \thead[r]{Line + Branch} & \thead[r]{Line} & \thead[r]{Branch} & \thead[r]{Line + Branch} \\
        \hline
        \coverup{}         & \PCT{63.8} & \PCT{49.2} & \PCT{60.3} & \PCT{82.4} & \PCT{66.7} & \PCT{79.6} \\
        \codamosa~(codex)  & \pct{53.5} & \pct{33.6} & \pct{48.7} & \pct{62.9} & \pct{35.6} & \pct{55.1} \\
        \codamosa~(gpt4o)  & \pct{50.6} & \pct{28.6} & \pct{45.2} & \pct{54.2} & \pct{19.0} & \pct{47.2} \\
        \coverup~(ablated) & \pct{38.8} & \pct{19.0} & \pct{34.0} & \pct{43.8} & \pct{7.3}  & \pct{38.5} \\
        \hline
        \coverup{} \tiny on PY suite & \pct{97.0} & \pct{99.2} & \pct{97.1} & \pct{100.0} & \pct{100.0} & \pct{100.0} \\
        \hline
    \end{tabularx}
\end{table}

        \begin{table}[]
    \small
    \caption{\textbf{[RQ1] \coverup{} outperforms MuTAP's coverage on the MT suite}.
        \label{table:mutap-overall-coverage}
    }
    \begin{tabularx}{\columnwidth}{@{} X | r r r | r r r @{}}
        \hline
        \multicolumn{1}{l|}{\thead{\vspace{-5pt}}} & \multicolumn{3}{c|}{\thead{\vspace{-5pt}Overall Coverage}} & \multicolumn{3}{c}{\thead{\vspace{-5pt}Median Per-Module Coverage}} \\
        \thead[l]{Test Generator} & \thead[r]{Line} & \thead[r]{Branch} & \thead[r]{Line + Branch} & \thead[r]{Line} & \thead[r]{Branch} & \thead[r]{Line + Branch} \\
        \hline
        \coverup{}              & \PCT{89.2} & \PCT{88.7} & \PCT{89.0} & \PCT{100.0} & \PCT{100.0} & \PCT{100.0} \\
        MuTAP~(codex few-shot)  & \pct{80.8} & \pct{71.7} & \pct{77.7} & \pct{100.0} & \pct{100.0} & \pct{100.0} \\
        MuTAP~(codex zero-shot) & \pct{79.2} & \pct{72.7} & \pct{77.0} & \pct{100.0} & \pct{100.0} & \pct{100.0} \\
        MuTAP~(gpt4o zero-shot) & \pct{78.0} & \pct{74.5} & \pct{76.8} & \pct{100.0} & \pct{100.0} & \pct{100.0} \\
        MuTAP~(gpt4o few-shot)  & \pct{15.9} & \pct{0.0}  & \pct{10.4} & \pct{16.7}  & \pct{0.0}   & \pct{11.1} \\
        \hline
    \end{tabularx}
\end{table}

    \fi
    
    \bibliographystyle{ACM-Reference-Format}
    \bibliography{emery,juan,coverup}

\end{document}